\pdfoutput=1
\documentclass[twocolumn,10pt]{article}
\usepackage[margin=0.5in]{geometry} 
\usepackage{amssymb}
\usepackage{amsthm}
\usepackage{amsmath}
\usepackage{amsfonts}
\usepackage{stmaryrd}
\usepackage{mathtools}
\usepackage{dsfont}
\usepackage{multirow}
\usepackage{adjustbox}
\usepackage{cite}
\usepackage{color}
\usepackage{hyperref}

\usepackage[flushleft]{threeparttable}
\usepackage{booktabs,caption,fixltx2e}
\usepackage{adjustbox}
\usepackage{bbding}

\usepackage{setspace}
\usepackage[runin]{abstract}
\usepackage{epigraph}
\makeatletter
\def\@xfootnote[#1]{%
  \protected@xdef\@thefnmark{#1}%
  \@footnotemark\@footnotetext}
\makeatother

\DeclareFontFamily{OT1}{pzc}{}
\DeclareFontShape{OT1}{pzc}{m}{it}{<-> s * [1.10] pzcmi7t}{}
\DeclareMathAlphabet{\mathpzc}{OT1}{pzc}{m}{it}

\newcommand{\blue}[1]{\textcolor{blue}{#1}}

\newcommand{\bx}{\mathbf{x}}
\newcommand{\R}{\mathds{R}}

\newcommand{\RRR}{{\R^3}}
\newcommand{\conf}{\mathsf{C}}

\newcommand{\SE}[1]{\mathrm{SE}({#1})}

\theoremstyle{definition}
\newtheorem{examp}{Example}

\title{\textsf{\textbf{Haptic Assembly and Prototyping: An Expository Review}}\protect\footnotemark}

\author{Morad Behandish and Horea T. Ilie\c{s}
\\
    {\small Departments of Mechanical Engineering and Computer Science and Engineering, University of Connecticut, USA}
}

\date{\small Technical Report No. CDL-TR-16-04, April 25, 2016}

\begin{document}

\maketitle

\footnotetext{A shorter version of this article is under peer-review for journal publication. For the time being, please use the following for citation:
    \protect\\
    \protect\\
        \blue{Behandish, Morad and Ilie\c{s}, Horea T., 2016. ``Haptic Assembly and Prototyping: A Survey.'' Technical Report No. CDL-TR-16-04, University of Connecticut.}
    \protect\\
    \protect\\
    The content of this expository article can also be found in Chapters 1 and 2 of the following thesis (published in April 2016):
    \protect\\
    \protect\\
        \blue{Behandish, Morad, 2016. ``Geometric Energies for Haptic Assembly.'' Master's Thesis, University of Connecticut.}
    }

\noindent \hrule \vspace{5pt}
\begin{abstract}
    An important application of haptic technology to digital product development is in virtual prototyping (VP), part of which deals with interactive planning, simulation, and verification of assembly-related activities, collectively called virtual assembly (VA). In spite of numerous research and development efforts over the last two decades, the industrial adoption of haptic-assisted VP/VA has been slower than expected. Putting hardware limitations aside, the main roadblocks faced in software development can be traced to the lack of effective and efficient computational models of haptic feedback. Such models must 1) accommodate the inherent geometric complexities faced when assembling objects of arbitrary shape; and 2) conform to the computation time limitation imposed by the notorious frame rate requirements---namely, 1 kHz for haptic feedback compared to the more manageable 30$-$60 Hz for graphic rendering. The simultaneous fulfillment of these competing objectives is far from trivial.

    This survey presents some of the conceptual and computational challenges and opportunities as well as promising future directions in haptic-assisted VP/VA, with a focus on haptic assembly from a geometric modeling and spatial reasoning perspective. The main focus is on revisiting definitions and classifications of different methods used to handle the constrained multibody simulation in real-time, ranging from physics-based and geometry-based to hybrid and unified approaches using a variety of auxiliary computational devices to specify, impose, and solve assembly constraints. Particular attention is given to the newly developed `analytic methods' inherited from motion planning and protein docking that have shown great promise as an alternative paradigm to the more popular combinatorial methods.

    In addition, we provide the most complete bibliography to date of haptic-assisted VP/VA systems complemented by a discussion and comparison of their key features.
\end{abstract}

\vspace{5pt} \hrule \vspace{20pt}

\newpage

\tableofcontents

\newpage

\section{Introduction} \label{sec_intro}

\epigraph{
    ``Smartness in mechanical devices is often realized through interaction that enhances dumb algorithms so they become smart agents...
   Interactive systems are grounded in an external reality both more demanding and richer in behavior than the rule-based world of noninteractive algorithms.''
   }{Peter Wegner, 1997 \cite{Wegner1997}}

\subsection{Human-Computer Interaction} \label{sec_HCI}

The role of `interaction' in taking full advantage of computers to support scientific advancements and day-to-day activities is becoming more evident with the rapid growth of hardware and software technologies. Although digital computers have been constantly breaking the records of data storage and processing over the past 50 years, their embarrassingly persistent failure in performing tasks as natural and intuitive to humans as facial recognition, speech recognition, language translation, and similar semantic interpretation problems has raised important questions. Specifically, what are the intrinsic (i.e., theoretical) limitations of the digital computers (and underlying computing models) in automating manual tasks? If computers are great at certain tasks---e.g., storing and processing vast amounts of data in short amounts of time---while humans are better at others---e.g., feeling, learning, and experiencing through the five senses---how can we characterize and leverage those capabilities simultaneously?

Rather than hopelessly trying to entirely automate all tasks that were once entirely manual, is it feasible to realize the `best of both worlds' though effective human-computer interaction (HCI)?%
\footnote{Also called human-machine or man-machine interaction or interface in the literature.}

\subsubsection{Historical Perspective}

Since the birth of the modern theory of computing and the notion of a universal computer by Alan Turing in 1936 \cite{Turing1936}, there has been no doubt that digital computers can far exceed the human ability to crunch numbers. Ever since, digital computing has revolutionized scientific disciplines, breached engineering limits, and eventually redefined everyday life after the advent of `personal computers' (PC).

\paragraph{Artificial Intelligence (AI).}

In spite of the seemingly limitless opportunities of computer-aided automation, the progress in certain directions has been surprisingly difficult. One such example can be seen in the prospects of artificial intelligence (AI) in 1980s. After Turning's landmark paper in 1950 \cite{Turing1950} followed by the famous Dartmouth conference%
\footnote{The Dartmouth conference and summer project is widely considered as the `birth of AI', where the phrase `artificial intelligence' was coined and the famous Dartmouth proposal was asserted: ``Every aspect of learning or any other feature of intelligence can be so precisely described that a machine can be made to simulate it.''}
in 1956 \cite{McCarthy2006}, there was a surge of research on AI for about two decades. Although the community's optimism grew rapidly during those `golden years' of AI, funding in AI suffered a major setback as researchers began to face the difficulties intrinsic to AI problems, and the AI research went through what is now known as the `AI winter' \cite{Crevier1993}. The question of whether `strong AI'%
\footnote{John Searle defined strong AI as the claim that ``the appropriately programmed computer with the right inputs and outputs would thereby have a mind in exactly the same sense human beings have minds.'' \cite{Searle1998}}
can be realized at all (or is even desirable anymore) remains unanswered today.

\paragraph{Human-Computer Interaction (HCI).}

In a different line of research on theory of computing, Peter Wegner observed that interaction is a fundamentally richer phenomenon than what classical model of algorithms can capture \cite{Wegner1997}. He showed that Turing machines cannot model interactive systems \cite{Wegner1998} and proposed `interaction machines' as a more powerful and expressive computing paradigm to model features of interactive behavior such as time and dynamically evolving adversaries and oracles \cite{Wegner1998,Wegner2012}.%
\footnote{He compares algorithms (without interaction) to `sales contracts' that deliver an output in exchange for an input, while (interactive) objects are like `marriage contracts' that specify the behavior for all contingencies of interaction---i.e., ``in sickness and in health [over the lifetime of the object] till death do us part'' \cite{Wegner1997}.}

With some of the roadblocks being removed by the availability of increasingly more powerful computers and richer computing models, the AI research has began thriving once again. Regardless of whether strong AI and full automation will be made possible in future (or ever at all) and whether or not one day the computers will surpass human cognitive capabilities indiscriminately, an important set of current problems can be solved more effectively by exploiting a collaboration (rather than competition) of humans and computers.

\subsubsection{Multimodal Interactions}

Any interaction between humans and their environment, including interaction with computers, can be viewed as an information exchange involving a combination of signals received via our five senses---namely, sight, hearing, touch, taste, and smell.%
\footnote{Perhaps with the exception of direct brain-computer interfaces (BCI) made possible for extremely simple tasks using electroencephalographic (EEG) signals collected on the human scalp. Research in this areas is still in the stage of infancy and is limited to few real-world applications. For a recent survey of BCI developments, see \cite{Ferreira2013}.}
However, the interfaces that we currently use for this type of interaction are often described as an `information bottleneck' \cite{Educause2007}. The substitution of punchcards and printers that endured as the primary device for information exchange even into the PC era until mid 1980s with today's interactive combinations such as keyboards, mice, joysticks, display devices, and more recently, touchscreens, has been a significant progress. However, the extent of the interactive phenomena that can be replicated by these limited set of gadgets is far from being comparable with those of many other human interactions in social and natural environments. These interactions range from those involved in regular day-to-day tasks such as tying one's shoe laces to specialized skills such as a craftsman's pottery of an artifact or a musician's playing of an instrument.

\paragraph{Engaging Multiple Senses.}

Significant progress has been made in both research and commercial capacities in incorporating two of the five senses quite effectively. These two senses are, obviously, the sense of hearing engaged by digital audio and the sense of sight engaged by computer graphics technologies. Since the commercialization of the television in late 1940s, the developments in both visual and auditory categories have been driven substantially by the multibillion dollar entertainment (e.g., motion picture and gaming) industry---not to mention other domains such as medical and defense applications.

The next frontier in modeling and implementing human sensorimotor capabilities involves the addition of the sense of touch. Unlike the visual and auditory information that can be more easily replicated, i.e., interpreted, stored, and recreated by (projections of) geometric representations and electrical signals, respectively, it is not very clear what the primitive information units for modeling touch should be. For this and other practical reasons explained in Section \ref{sec_challenges}, the touch-enabled HCI has not seen the same rapid growth. The most notable examples of its successful deployment today are either in specific application domains with a restricted and well-understood role for touch---e.g., flight simulation, clinical training, rehabilitation, and alike---or in simplified scenarios that involve a narrow and primitive subset of sensory experience---e.g., touchscreen devices with vibration feedback. In these applications, the type of touch interaction is extremely restrained to facilitate designing special purpose devices that engage only certain parts of the body---e.g., through particular muscles/joints or only fingertips---or involve confined interaction modes---e.g., force feedback at a single point while holding a stylus with a natural grasp like holding a knob, wand, or stylus.%
\footnote{A good example application domain is in clinical training. On the one hand, there is robotic surgery where the interaction modes are quite diverse and challenging to formalize except in a restricted subset of tasks---e.g., cutting tissue using scissors with 1 DOF. On the other hand, there is dental diagnosis which involves little more than what a haptic stylus with a single-point 3 DOF force feedback can replicate. As a result, one observes more mature haptic hardware/software in the latter category currently on the market for training purposes.}
The reader is referred to \cite{Laycock2003,Hayward2004,Tsalamlal2013} for examples of haptic devices on the market.

\begin{figure*}
    \centering
    \includegraphics[width=0.75\textwidth]{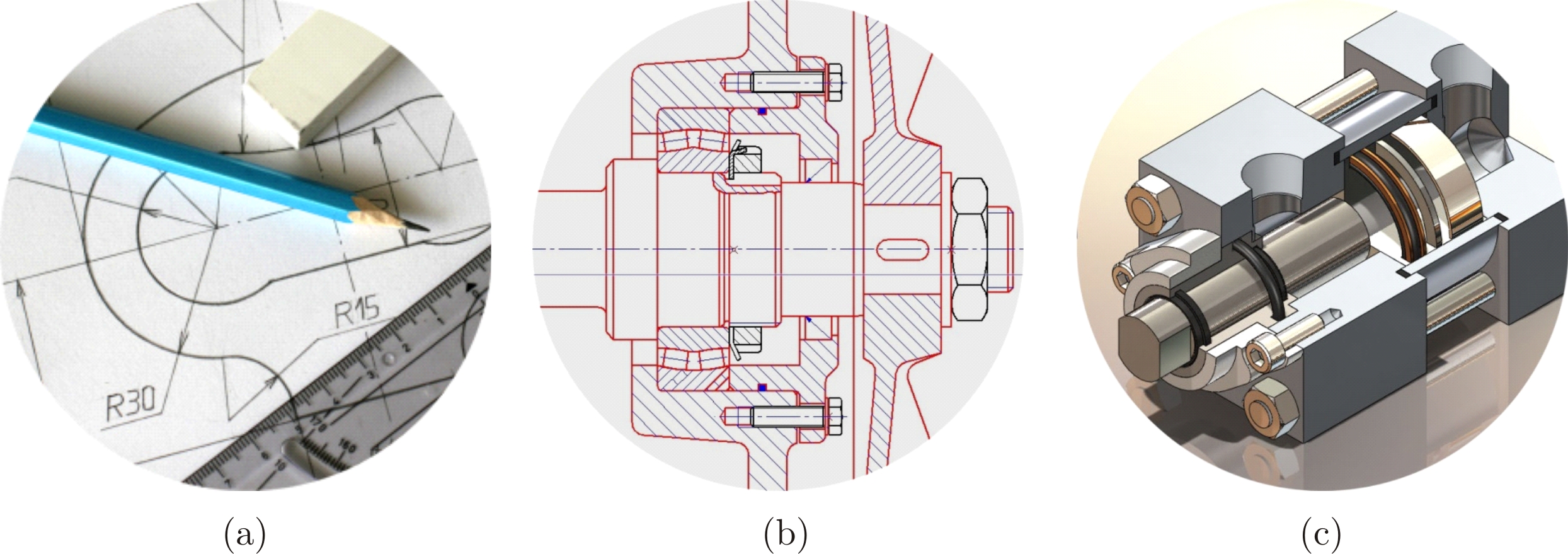}
    \caption{The evolution of visual aids for mechanical design; from paper-and-pencil sketching (a) to computer-aided tools from 2D drafting (b) to 3D modeling (c).} \label{figure1}
\end{figure*}

\subsubsection{User Interfaces for CAD}

To understand the importance of haptic interfaces, in general, and the implications of developing foundations and algorithms to support touch-assisted engineering tools, in particular, one needs to appreciate the analogous role that the graphical user interface (GUI) has played in the evolution of computer-aided design, analysis, and manufacturing.

\paragraph{Graphical User Interface (GUI).}

It is hard to underestimate the crucial role of visual aids (e.g., symbols, sketches, and diagrams) in carrying out even the simplest cognitive processes%
\footnote{``Two minutes with a pencil on the back of an envelope lets us solve problems which we could not do in our heads if we tried for a hundred years.'' --Christopher Alexander \cite{Alexander1964}.}
let alone its importance in scientific and engineering practice. Architects, engineers, and artists have used sketches for hundreds of years to conceptualize, examine, and communicate their creations. The advent of digital computing and its democratization during the 20th century provided the designers with an incredibly powerful new set of tools. However, to take full advantage of the computational power, there was no other choice but to equip this new machine with visual aids to replace the traditional sketch pad, paper-and-pencil in assisting the designer's imagination (Fig. \ref{figure1}).

\begin{figure*}
    \centering
    \includegraphics[width=\textwidth]{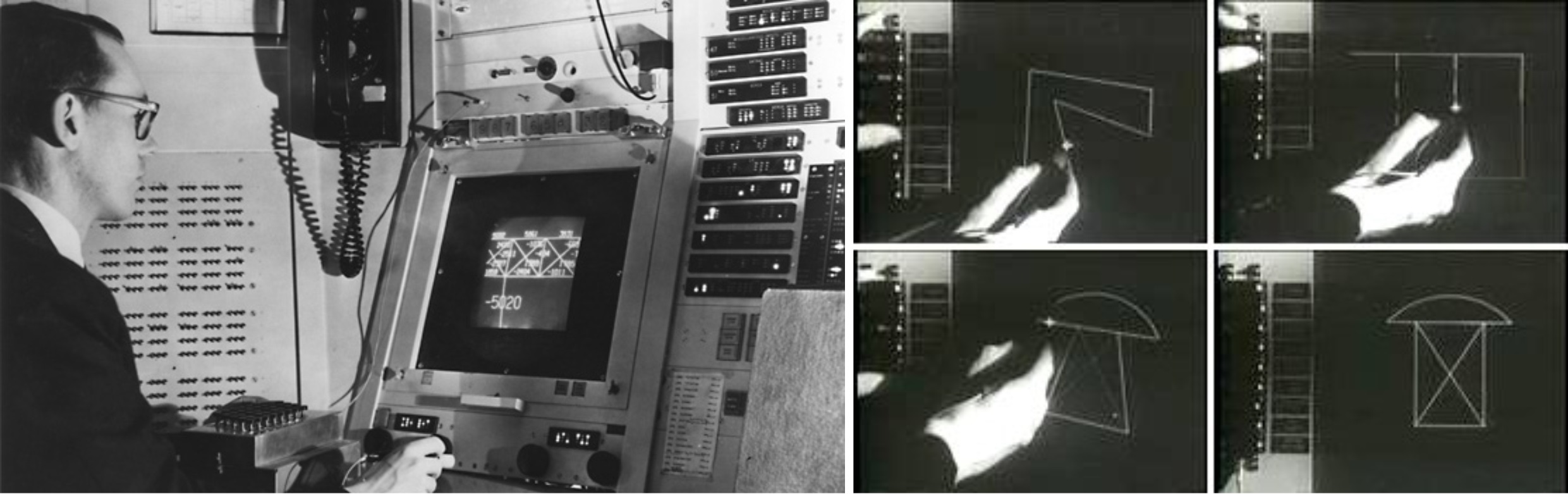}
    \caption{Ivan Sutherland showcasing \textsf{Sketchpad} at MIT's Lincoln Labs.} \label{figure2}
\end{figure*}

In 1963, Ivan Sutherland developed a revolutionary computer program called \textsf{Sketchpad}: ``a man-machine graphical communication system'' \cite{Sutherland1963} (Fig. \ref{figure2}). \textsf{Sketchpad} can be viewed as the first computer-aided design (CAD) software and a major breakthrough in the development of CAD and computer graphics as disciplines. It gave birth to the concept of a GUI and spurred a new approach to HCI.

In 1972, Herbert Voelcker and his collaborators started the production automation project (PAP) \cite{Voelcker1993} aiming to develop the solid modeler called \textsf{part and assembly description language (PADL)} \cite{Staff1974}. PAP was originally established aiming to provide informationally complete computer representations to support the automation of numerical control (NC) machining. \textsf{PADL-1} \cite{Fisher1977} used a combination of constructive solid geometry (CSG) and boundary representation (B-rep) and was made publicly available in 1977. \textsf{PADL-2} \cite{Requicha1974a} followed closely after in 1981 from which \textsf{UniSolid} was developed by \textsf{Unigraphics}$^\textrm{TM}$. Around the same time in 1978 the B-rep solid modeling kernel \textsf{Romulus} was released by \textsf{ShapeData}, which influenced the successor kernels \textsf{Parasolid}$^\circledR$ and \textsf{ACIS} in late 1980s and early 1990s. Other notable CAD software that emerged in those years are \textsf{AutoCAD}$^\circledR$, \textsf{CATIA}$^\circledR$, and \textsf{Pro/ENGINEER}$^\circledR$ in 1980s and \textsf{SolidWorks}$^\circledR$, \textsf{SolidEdge}$^\circledR$, and \textsf{Autodesk Inventor}$^\text{TM}$ in 1990s. See \cite{Carlson0000} for a broad historical review of the evolution of computer graphics and CAD industry, and \cite{Weisberg2008} for a more detailed review of early CAD packages in particular.

In the past two decades, the 3D modeling, simulation, and visualization tools for CAD have evolved into versatile product lifecycle management (PLM) software to manage complete digital product descriptions for development, design, and manufacturing known as digital mock-ups (DMU) \cite{Bullinger1999}. DMU serves as a platform for virtual (or digital) prototyping (VP/DP), which is becoming a commonly adopted design and validation practice in several industrial sectors \cite{Bordegoni2006}. Although today's versatile GUIs allow an interactive visual examination of 3D models at different design and validation stages, a major deficiency can be traced to handling 3D geometry and functions using 2D input/output media \cite{Bullinger1999}. Almost all current CAD software use a combination of widget-based interaction using a pointing device (e.g., a mouse or touchpad) cursor on the screen---namely, the standard windows, icons, menus, and pointer (WIMP) interaction style---and alphanumeric input using a keyboard \cite{Bourdot2010}, while research has shown that multimodal interactions enable higher accuracy and lower execution times \cite{Hecht2006}.

\paragraph{Haptic User Interface (HUI).}
Several general-purpose robotic devices meant for touch-enabled input/output interactions with computers (called `haptic interfaces') ranging from 3D to 6D input (i.e., position and orientation encoding) and output (i.e., force and torque feedback) became widely available in 1990s. Among the most widely used commercial devices one can refer to \textsf{SensAble}$^\circledR$ \textsf{Phantom}$^\circledR$ line of devices that originated from MIT Touch Lab \cite{Massie1996,Salisbury1997} (now called \textsf{Geomagic}$^\circledR$ \textsf{Touch$^\mathrm{TM}$} upon acquisitions in 2012 and 2013), \textsf{Haption Virtuose6D}$^\text{TM}$ popular for assembly planning with large workspace requirements, \textsf{CyberGlove}$^\circledR$ \textsf{CyberGrasp}$^\circledR$, \textsf{Force Dimension} \textsf{Sigma}/\textsf{Omega}/\textsf{Delta}, and \textsf{Novint Falcon}.%
\footnote{We hereby emphasize that exemplifying or mentioning device brands that have been used for testing purposes in the reviewed technical papers do not imply recommendation of these devices over any others by the authors. See \href{www.bracina.com/haptichardware.html}{www.bracina.com/haptichardware.html} for a list of haptic devices on the market.}
Although these general-purpose devices have found numerous applications in research and development labs, as well as special-purpose devices developed in-house and optimized for certain applications, their democratization for the consumer market has been hindered by
\begin{enumerate}
    \item the high hardware retail prices due to limited number of units produced;
    \item the lack of common language standards and general-purpose software development platforms to make the technology accessible to non-experts; and
    \item the absence of a clear customer demand---which is both responsible for and aggravated by the above two factors.
\end{enumerate}
The reader is referred to \cite{Laycock2003,Hayward2004,Tsalamlal2013} for surveys of the evolution of haptic interfaces and to \cite{Srinivasan1997,ElSaddik2011,Varalakshmi2012} for the extent of their applications in engineering, medicine, entertainment, and education.

We postpone a more detailed review of the application of haptics to assist VP/VA to Section \ref{sec_CADapps}. Here it suffices to emphasize that in a similar fashion that developments in `computer graphics' have been essential to achieve the current functional versatility of interactive 3D modeling tools for CAD, the integration of haptic interfaces to achieve their full potential calls for extensive research and development in `computer haptics'.%
\footnote{Haptic studies are typically organized into three subareas; namely, human haptics, machine haptics, and computer haptics \cite{Srinivasan1995}, which study the physiological, hardware, and software aspects, respectively. Being particularly relevant to this article, computer haptics \cite{Ho2000} is defined as algorithms and software associated with integrating {touch} into (HCI), analogous to computer graphics targeting {sight} \cite{Srinivasan1995}.}
Developments in both graphics and haptics interaction modes are, in turn, dependent on the advancement of more inclusive mathematical abstractions (e.g., geometric modeling and reasoning) and richer computer representations (e.g., data structures and algorithms).

\subsubsection{Practical Significance} \label{sec_significance}

Among the numerous applications of graphics- and haptics-enabled HCI, haptic feedback is of particular practical significance in application domains that involve an interplay of shapes and motions. More specifically, when dealing with planning problems that require complex spatial reasoning in higher-dimensional configuration spaces---e.g., assembly planning, path and motion planning, manufacturing process planning, etc.---there are two extreme ends to a computer-aided solution process:
\begin{itemize}
    \item The algorithmic approach involves attempting to model a `field description' of the problem \cite{Alexander1964}---e.g., define a metric to quantify the relative performance of the different plans---followed by devising a search algorithm to find a solution---e.g., an optimization algorithm that navigates through the configuration space and finds local and/or (idealistically) global optima by trial-and-error.
    \item The interactive approach involves leveraging the human agent's expertise and domain knowledge to reject a large subset of trials upfront and eventually select an acceptable (if not optimal) solution by interactive evaluation---e.g., visual and sensorimotor inspection of the configuration space via graphic and haptic feedback, respectively.
\end{itemize}
Clearly, each of the above approaches has its own advantages and drawbacks. For noninteractive optimization, one needs to come up with a purely quantitative metric to evaluate each and every candidate plan as accurately as possible, which is unrealistic when inevitable heuristics with limited applicability are involved. Even if such a uniform field description is discovered, finding the near-optimal solution requires searching a high-dimensional solution space which is computationally prohibitive, especially in the presence of constraints.%
\footnote{Mathematically, a highly constrained search problem---without the knowledge of a parameterization that explicitly guarantees constraint satisfaction---is an attempt to find lower-dimensional manifolds in a higher-dimensional space, which depicts the computational challenge faced in the algorithmic approach by appealing to simple probabilistic arguments.}
Most search algorithms are doomed to converge to local optima without any global guarantees, and are generally unable to systematically partition the search space to what can be referred to as ``qualitatively distinct'' solution subspaces or regions with few plausible recent developments \cite{Nelaturi2014,Nelaturi2015b}. The interactive approach, on the other hand, allows the human agent to identify qualitative distinctions and limit the search space to a few regions while rejecting the rest. The quantitative field description can then assist the user---through proper visual and sensorimotor cues---in finding the local minima in each subspace and further refining the solution.

Besides interactive assembly planning at the focal point of this paper, other related spatial planning problems can also benefit from an energy field description. In particular, automatic robot path planners (including but not limited to assembly planners) that use decomposition methods such as cylindrical algebraic decomposition (CAD$^\ast$) \cite{Schwartz1983,Schwartz1983a,Schwartz1983b} or faster sampling methods such as probabilistic roadmaps (PRM) \cite{Kavraki1998,Kavraki1996,Boor1999,Geraerts2004} to construct a combinatorial structure over the configuration space can skip the redundant exploration of a large subset of infeasible solutions by restricting the search space to that swept by the user through interaction. Similarly, protein docking platforms that rely on FFT-based search algorithms \cite{Katchalski1992,Chen2003a,Kovacs2003,Eisenstein2004,Bajaj2011} to explore the configuration space can exploit the user's ability to quickly identify qualitative docking sites before making quantitative adjustments to the local configurations.

In the next section, we briefly enumerate the current challenges and promising directions for incorporating haptics into assembly planning applications.

\subsection{Research Challenges} \label{sec_challenges}

The growth in the availability and popularity of the fairly recent haptic technology imposes increasing demands for geometric modeling and computing algorithms, to deliver realistic replication of the real-world experience in virtual environments (VE) as efficiently as possible.
We identify the following main challenges in the development of haptic-enabled virtual prototyping tools---with virtual assembly tasks being considered as the particular subarea of interest:
\begin{enumerate}
    \item capturing the inherent geometric and spatial complexities faced when assembling objects of arbitrary shape; particularly, the absence of a practically effective force and torque feedback model for assembly and disassembly interactions of general mating pairs; and
    \item abiding by the computation time limitation imposed by the frame rate requirements; particularly, the obligation to carry out all computations per frame within a millisecond to maintain a 1 kHz servo-loop rate for haptic feedback compared to the more manageable 30$-$60 Hz for graphic rendering.
\end{enumerate}
The bigger difficulty is to address both of the above challenges {\it simultaneously}. The above two aspects are described in more detail in the following Sections \ref{sec_geomcomp} and \ref{sec_complimit}, respectively.

\subsubsection{Geometric Complexities} \label{sec_geomcomp}

The primary challenge in developing generic models for haptic feedback lies in a proper formulation of the guidance forces and torques that effectively assist the user in the exploration of the VE, from repulsing collisions to attracting proper contact. In particular, when two objects (e.g., rigid or flexible parts or subassemblies) are being assembled or disassembled in a VE, the effects of the mechanical forces and torques exchanged during the process are simulated by integrating the equations of motion subjected to the constraints formulated using contact mechanics and friction models.
As depicted in Section \ref{sec_PBM}, for the most general type of contact---i.e., a combination of surface, curve, and point contact---collision detection (CD) algorithms are used for computing the resistance forces and torques that are central to the dynamic simulation. However, there are several computational challenges faced when attempting to perform CD in real-time interactive applications. This is especially the case in assembly scenarios that involve tight fits whose `nominal' geometry is described by contact features that reduce the degrees of freedom (DOF) of relative motion---e.g., by restricting the 6D relative translations and rotations of a pair of rigid bodies into lower-dimensional (e.g., 1D or 2D) subspaces of their configuration space. The dynamic simulation along the lower-dimensional contact subspaces for nominal geometry is inherently unstable and difficult to compute due to the extreme sensitivity to infinitesimal perturbations. The natural solution, similar to the case of real physical assembly, is to add finite {\it clearances} according to appropriate geometric dimensioning and tolerancing (GD\&T) standards \cite{Srinivasan2008} to the nominal geometry for ease of assembly:%
\footnote{Here we are specifically considering the virtual simulation of the insertion for `clearance fits' that the user can do with bare hands (in both physical and virtual setups alike) and do not require special tools or processes unlike the instances with `force fits', `shrink fits', etc.}
\begin{itemize}
    \item High-clearance fits have been successfully simulated in VEs---e.g., clearances of $\sim$1$-$3 mm using approximate CD with resolution of $\sim$0.2 mm \cite{Seth2006,Seth2008}. However, most real assembly scenarios require much smaller clearances in accordance with GD\&T specifications \cite{ASME2009}.
    \item Low-clearance fits, on the other hand, require more accurate CD algorithms---e.g., clearances of $\sim$0.001 mm using exact CD on original B-reps \cite{Seth2007,Seth2010}. However, exact CD does not exhibit the required performance at 1 kHz with the existing hardware capabilities.
\end{itemize}
In other words, there is an inevitable trade-off: the practical and efficient CD methods use some approximation---e.g., meshing, voxelization, or bounding volume hierarchies---that compromises the required accuracy for low-clearance assembly, while exact CD is not fast enough to handle numerous parts or complex shapes in real-time. The different CD methods for haptic assembly are reviewed in Section \ref{sec_CD} and in greater detail for general applications in \cite{Lin1998,Jimenez2001,Kockara2007}.

\paragraph{Sources of Geometric Errors.}
It is important to note that the dynamic instability problem of low-clearance fits whose nominal geometry typically characterizes a lower kinematic pair---i.e., contact maintained over a surface, restricting the motion to 1D or 2D---is actually twofold when it comes to interactive VR applications:
\begin{enumerate}
    \item the intrinsic geometric representation errors of the approximate CD algorithm---e.g., voxelized or triangulated cylindrical features---which can be eliminated in principle by using exact CD at the expense of computational performance; and
    \item the measurement errors/noise due to the haptic device encoder inaccuracies as well as the `jerking' motion of the hand, which can only be alleviated (e.g., by filtering) at the software level but cannot be completely eliminated.
\end{enumerate}
Unlike what happens in real physical assembly, the dynamic behavior in virtual assembly is simulated using a finite-difference integrator, whose stability is very sensitive to these errors. As a result, CD alone has been found by several researchers \cite{Vance2011,Perret2013} to be insufficient for virtual assembly, especially with haptic assistance.

\paragraph{Recognizing Mating Features.}

For the lower kinematic pairs that are completely classified into the well-known six classes \cite{OConnor1996} (detailed in Section \ref{sec_motion}), it is possible to abstract the DOF-limiting contact subspaces in terms of `virtual fixtures' \cite{Rosenberg1993} or `mating constraints' \cite{Iacob2011} between functional surfaces of simple (e.g., planar, cylindrical, spherical, or conical) shapes. The forces and torques for haptic guidance during insertion of these features are then simplified, for example, by using spring-damper models between the current and eventual configurations. Although such approaches provide faster and more effective alternatives to CD for low-clearance assembly, they also depend on at least one of the following simplifications:
\begin{itemize}
    \item a priori assumptions on the type of surface features and corresponding kinematic pairs, their explicit semantics, and exact locations on the different parts and subassemblies {\it manually} specified either by the CAD designer (pre-importing) or by the VE user (on-the-fly); or
    \item heuristic methods to {\it automatically} detect the assembly intent and associated mating features as soon as the features are brought to insertion proximity.
\end{itemize}
The abstraction of the mating features for higher kinematic pairs between features of arbitrary shape, on the other hand, is significantly more difficult since no such finite classification exists for the more general case.

As a consequence of these disadvantages in using each method by itself, the common theme in the recent haptic assembly systems is to use hybrid techniques \cite{Vance2011} (detailed in Section \ref{sec_hybrid}) that switch between CD and feature-based constraint resolution. This duality creates extra complications for switching and blending between the two phases \cite{Perret2013}. Moreover, when fixated on the low-clearance fit of a particular pair of features, collision events outside the insertion site can be missed. This defeats one of the main purposes of virtual assembly, which is the early detection of design issues such as unaccounted clashes between the different parts.

\subsubsection{Computational Limits} \label{sec_complimit}

The efficiency problem appears more challenging in the case of haptic feedback, when compared to graphic rendering, due to the notorious physiological requirement of (at least) 1 kHz refresh rate necessary for satisfactory tactile experience---especially to acquire the necessary stiffness when manipulating rigid objects \cite{Perret2013}---while only 30$-$60 Hz is typically perceived as adequate for appealing to human vision \cite{ElSaddik2011}.

\paragraph{Frame Rate Requirements.}
The human touch perception is typically classified into `kinaesthetic' sensations that are related to muscle control and limb motion, and `tactile' sensations perceived at the skin receptors \cite{ElSaddik2011}.%
\footnote{More technically, the term `proprioception' is used for the broad class of perceptions of the position, state, and movement of the body and limbs in space. This includes vestibular, kinaesthetic, and cutaneous sensations. The `vestibular' sensations pertain to the perception of balance, head position, and acceleration/deceleration. The `kinaesthesia' includes the sensation of movement of the body and limbs originating in the muscles, tendons, and joints. The `cutaneous' sensations pertain to the skin itself, including sensations of pressure (from mechanoreceptors) as well as temperature and pain (from nociceptors) the former of which is more specifically referred to as `tactile' \cite{Paterson2007}.}
Although a bandwidth of 10 Hz is typically considered adequate for the kinaesthetic sensations \cite{Srinivasan1995}, it is largely dependent on the task, e.g., 1$-$2 Hz for unexpected signals, 2$-$5 Hz for periodic signals, up to 5 Hz for internally generated or learned trajectories, and 10 Hz for reflex actions \cite{Srinivasan1995,Srinivasan1997}. However, to adequately simulate rigidity with a force feedback device (e.g., in VR-CAD applications), higher frequencies are required due to basic control-theoretic considerations; namely, noting that the maximum stiffness in a closed-loop system is inversely proportional to the square of the regulation period \cite{Perret2013}. On the other hand, vibrations of up to 1 kHz can be resolved by the human tactile system, with the highest sensitivity at 250 Hz \cite{Srinivasan1997}. To collectively comply with all of these requirements, a response rate requirement of 1 kHz is widely accepted as the standard for most haptic applications---see \cite{Burdea1996,Srinivasan1995,Srinivasan1997,ElSaddik2011} for more details on the subject. It should be noted, however, that there are studies suggesting that even higher rates of 5$-$10 kHz might be necessary for improved haptic performance in certain tasks \cite{Kubus2009} one example being high-fidelity texture discrimination \cite{Choi2004,Choi2005}.

\section{Literature Review} \label{sec_lit}

The rest of this paper is devoted to a survey of research and development in haptic-assisted VP/VA in Section \ref{sec_CADapps}, a classification of the dominant existing themes for assembly constraint modeling in Sections \ref{sec_PBM} and \ref{sec_CBM}, and a proposal for future directions and opportunities in Section \ref{sec_future}.

\subsection{CAD/CAAP Applications} \label{sec_CADapps}

Today, most engineering design tasks are heavily assisted by powerful and widely available computer simulation and visualization tools. Although a large subset of analysis and synthesis tasks have been partially (if not fully) automated, the designer's decision-making remains central to certain aspects of the design process. This in turn creates a demand for more effective human-computer interfaces to explore more efficient, creative, and cost-effective design solutions in semi-automatic setups.

In particular, haptic assistance has been found useful in several design activities that can benefit from domain expertise and cognitive capabilities of human operators---which are hard to formalize for full automation such as conceptual design \cite{Bordegoni2006,Bordegoni2006b,Bordegoni2010}, aesthetic design \cite{Bordegoni2008,Bordegoni2010a,Bordegoni2012}, design review and functionality validation \cite{Moreau2004,Yang2005,Ferrise2010,Araujo2010}, ergonomics and human factors evaluation \cite{Shaikh2004,Jayaram2006,Bordegoni2007a,Bordegoni2007}, and many more reviewed in Appendix \ref{app_VP}. In particular, haptic manipulation has also been leveraged for editing parametric CAD models and freeform surfaces for designing individual parts \cite{Dachille2001,Dachille1999,Liu2004a,Liu2004,Liu2005,Gao2005,Gao2005a,Gao2006}. Particular attention will be given to methods concerned with applying haptics to the evaluation and planning of assembling and disassembling (already designed) rigid and flexible parts---another critical and costly step in the product design process \cite{Boothroyd1987}.

A thorough review of research efforts and organization of the literature on graphics- and haptics-assisted virtual assembly can be frustrating, as appreciating its current position and potential implications in the modern product life-cycle management (PLM) requires an overview of a range of different topics. In this section, we provide a sufficiently broad review---brief in text, less so in the number of citations---of some of the relevant research studies in applying virtual reality (VR) and augmented reality (AR) tools to computer-aided design and assembly planning (CAD/CAAP), virtual prototyping (VP) (Section \ref{sec_VP}), and virtual assembly (VA) (Section \ref{sec_VA}). Of course we do not intend, by any means, to provide a complete survey of VP/VA related research or all published industrial implementations. Instead, we shall focus on providing the reader with clear {\it definitions} and {\it classifications} of the existing approaches to addressing the difficulties described in Section \ref{sec_challenges} along with introducing common terminologies, comparing advantages and drawbacks, and citing (more than enough) pointers if further details are sought. Particular attention is given to the different conceptualizations of {\it constraints} (Section \ref{sec_motion}) that emerge in VR-CAD assembly and disassembly problems. Constraint modeling and resolution are central to solving the motion of arbitrarily complex parts in contact in real-time---attempting to simultaneously address both of the challenges presented in Section \ref{sec_challenges}---using physics-based (Section \ref{sec_PBM}), constraint-based (Section \ref{sec_CBM}), or combined (Section \ref{sec_future}) techniques.

\subsubsection{Virtual Prototyping} \label{sec_VP}

Recently, an early-stage examination of different product life-cycle aspects related to design, manufacturing, maintenance, service, and recycling has been made possible by integrating VR tools into the modern CAD environments, a practice referred to as virtual (or digital) prototyping (VP/DP) \cite{Gupta1997,Bullinger1999,Wang2002,Deviprasad2003}. Such an evaluation results in a significant reduction of time and cost associated with physical prototyping (PP), and allows for the elimination of a large subset of design issues in the earlier stages of the process \cite{Coutee2002,Coutee2001}. Although they cannot yet completely replace physical prototypes, virtual prototypes are less expensive, more repeatable, and easily configurable for different variants, hence provide significant insight into the functionality of the product while eliminating redundant design trials and excessive tests \cite{Bordegoni2006}.

In the larger context of modern PLM, one often encounters the notion of a digital mock-up (DMU). DMU consists of a complete digital descriptions of the product during its entire life-cycle, serves as a platform for product and process development, and includes geometric, ergonomic, and functional information---with or without human-computer interaction element. DMU construction accounts to realistic computer simulations that are capable of replicating different functionalities ranging from design, manufacturing, maintenance, service, and recycling of the physical mock-up (PMU) \cite{Gomes1999}. The interactive application of immersive VR tools for a subset of those tasks, including (but not limited to) assembly and disassembly process verification, ergonomics and functional assessment, and other design evaluations are the subset of DMU development technologies referred to as VP \cite{Gomes1999}.%
\footnote{The notions of physical/digital mock-ups (PMU/DMU) and physcial/digital prototyping (PP/DP) are sometimes used interchangeably, but it is safe to say that a more precise definition puts the latter as a subset of the former that involves VR technologies \cite{Gomes1999}.}

For reviews of PP versus VP techniques and their classifications, capabilities, and limitations for product development, we refer the reader to \cite{Bernard2002,Zorriassatine2003,Gibson2004}. Currently, the most notable industrial applications of VP are found in the automotive and aeronautic industries \cite{Dai1996,Bennett1997,Zimmermann2008}. Instructive (although not very up-to-date) surveys of manufacturing applications in general can be found in \cite{Shukla1996,Moreau2004,Mujber2004}. In particular, an assessment of the capabilities of VR hardware and software tools available in early 2000s to support VR integration into product life-cycle management (PLM) is given by Jayaram et al. \cite{Jayaram2001}.

Appendix \ref{app_VP} gives a detailed account of several important and relatively recent studies and systems that use haptic support for a variety of design, analysis, validation, and manufacture process planning activities. Table \ref{tab_VPlit} provides a more extensive and chronological list of studies along with their hardware and software components and key features.

\subsubsection{Virtual Assembling} \label{sec_VA}

\begin{figure}
    \centering
    \includegraphics[width=0.48\textwidth]{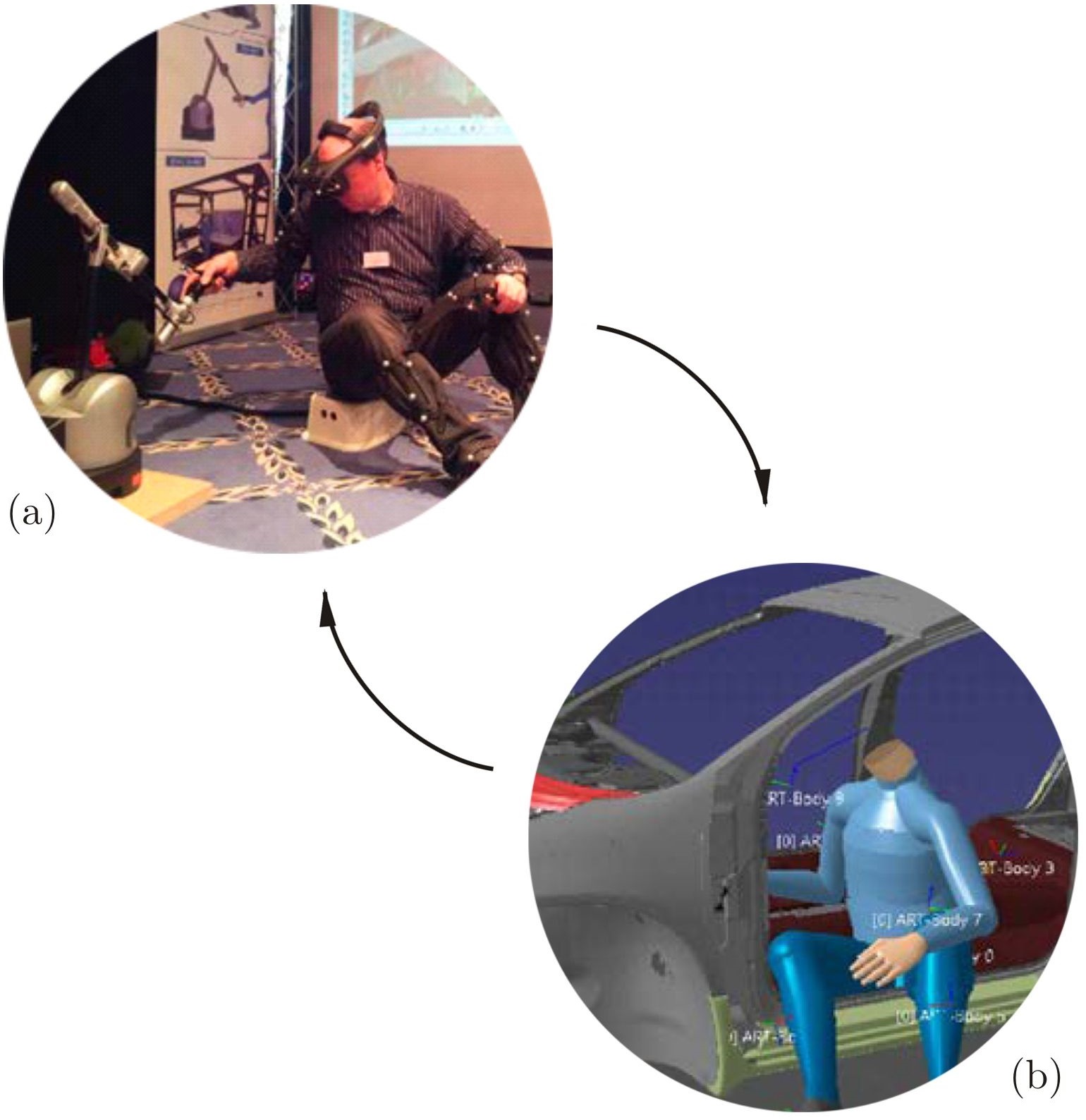}
    \caption{Application of haptic technology to virtual assembly and prototyping. Pictures courtesy of Perret et al. \cite{Perret2013}. } \label{figure3}
\end{figure}

Among other VP activities, virtual assembly (VA), defined as a simulated assembly of the virtual representations of mechanical parts in an immersive 3D user interface using natural human motions \cite{Kim2003,Kim2004} (Fig. \ref{figure3}), characterizes an important subarea of VP, to which applying haptic feedback has been shown particularly beneficial in terms of task efficiency and user satisfaction by multiple researchers \cite{Gomes1999,Volkov2001,Lim2007,Wildenbeest2013}. For instance, a user survey carried out in \textsf{BMW} by Gomes and Zachmann \cite{Gomes1999} predicted an important role for VR tools in prototyping and assembling activities in the future of the automotive industry. In particular, they showed that VR-enabled DMUs reduce the need for PMUs and facilitate an improvement of the overall product quality. However, the study concluded that a widespread use of VR in manufacturing industries is contingent upon its seamless and complete integration into CAD/CAAP. Volkov and Vance \cite{Volkov2001} showed that haptic assistance improves the ability of a user for design decision making in VP/VA environments; particularly in terms of task efficiency (e.g., less time taken for user evaluation) and user satisfaction when evaluating automotive design examples. Lim et al. \cite{Lim2007} showed that small (i.e., visually insignificant) geometric features---e.g., chamfer or fillets in a simple peg-in-hole pair---can significantly affect user performance in haptic assembly, with measurements showing a similar trend to those of physical assembly. Their results demonstrate that adding haptic feedback to the VP/VA process enables exploiting shape information that are significantly underused when only visual feedback is provided. Wildenbeest et al. \cite{Wildenbeest2013} conducted experiments to investigate the impact of haptic feedback quality in the performance of teleoperated assembly in the context of four sub-tasks; namely, free-space movement, contact transition, constrained translational, and constrained rotational tasks. They concluded that low-frequency haptic feedback improves overall task performance and control effort in constrained translational and rotational tasks.

In the past two decades, there have been numerous studies and systems focused on the development of immersive virtual environments for solving assembly and disassembly problems. These systems have used a variety of visualization tools (e.g., stereoscopic displays and goggles) and tracking devices (e.g., head tracking devices and data gloves) to assist the user in virtual object manipulation tasks. More recently, an increasing number of studies have leveraged haptic devices to provide a more realistic assembly experience with force feedback, a thorough account of which would constitute enough material for a full book on the subject. Different survey articles have been authored recently, each focusing on different aspects of the haptic assembly technology:
\begin{itemize}
    \item See Seth et al. \cite{Seth2011} for an earlier qualitative review of previous studies and existing systems on VP/VA with and without haptic feedback. Here we focus more on the haptic-assisted systems in greater detail, also including the more recent developments.
    \item See Xia et al. \cite{Xia2013} for a brief overview of the existing challenges and research directions in VR and particularly in leveraging haptics for product assembly with rigid parts and soft cables. Here we intend to identify and analyze the most important challenges in greater depth and propose a precise research agenda.
    \item See Leu et al. \cite{Leu2013} for a state-of-the-art report on the methods for CAD model generation from digital data acquisition, motion capture, assembly modeling, HCI, and CAD-VR data exchange, to enable assembly simulation, planning, and training. Their report covers different (and important) aspects of CAD-VR integration that we shall not discuss here.
    \item See Vance and Dumont \cite{Vance2011} for a popular proposal for a promising future direction in haptic assembly---to which we will return in Section \ref{sec_hybrid}.
    \item See Perret et al. \cite{Perret2013} for a discussion of some of the technical issues faced by haptic assembly in practice along with an assessment of the technical maturity using the technology readiness level (TRL) index---to which we will return in Section \ref{sec_readiness}.
\end{itemize}

Other decent reviews on the subject from multiple points of view---followed by presentations of specific systems or methods---can be found in \cite{Lim2010,Iacob2012}.

This survey distinguishes itself by presenting conceptual insight into the back-end abstractions of geometric features that lead to formulating assembly constraints and computing guidance forces and torques for haptic feedback. Although the main theme is of geometric modeling and spatial reasoning techniques, we also provide the most complete catalogue of previous studies and systems to date, which could serve as a useful collection and a comprehensive ``beginner's guide'' to the literature.

Appendix \ref{app_VA} gives a detailed account of several important and relatively recent studies and systems that use haptic support for assembly and disassembly related problems among other prototyping activities. Table \ref{tab_VAlit} provides a more extensive and chronological list of studies along with their hardware and software components and key features.

\paragraph{Computational Commonalities.}

As catalogued in the appendices, there is a bewildering variety of VR systems implemented using different software libraries and computational tools for haptic-assisted assembly and disassembly activities. In spite of their major differences in terms of setup (e.g., desktop versus \textsf{CAVE}-like systems), architecture (e.g., PC versus network-based systems), and hardware (wand-type versus glove-type devices), the underlying geometric and physical modeling of most systems are quite similar; namely, they all use the following basic tools:
\begin{itemize}
    \item collision detection (CD) to avoid penetration between parts and subassemblies, typically available as part of the physics simulation engines (PSE), which can be conceptualized in terms of `physical constraints';
    \item heuristic attraction and/or repulsion forces and torques formulated using spring-damper models or some other simple potential energy field derived from CAD model's mating pairs called `geometric constraints';
\end{itemize}
or a blended combination of both. But as we shall articulate in the rest of this section, CD-based physical constraints are difficult to resolve in the vicinity of tight (i.e., low-clearance) fits for a variety of reasons (see Section \ref{sec_PBM}), while the artificial geometric constraints are too simplistic to handle arbitrary geometry (see Section \ref{sec_CBM}). The following section clarifies the conceptual distinctions between different types of constraints and their implications for VR implementations.

\subsubsection{Constrained Motion} \label{sec_motion}

The simulation of assembly and disassembly processes for rigid parts can be abstracted as a free motion with 6 DOF (3 for translations and 3 for rotations) per part along with a set of {\it constraints} that restrict motion along those DOFs and create interdependencies across different parts. The existing approaches for simulating `part behaviour' in VR systems are typically classified into two groups, with the following definitions \cite{Badillo2013}:
\begin{itemize}
    \item physically-based modelling (PBM), which uses Newtonian/Lagrangian dynamics to solve for the motion trajectories of the virtual objects, under the effect of forces and torques due to physical contact between those objects (e.g., no-penetration impact forces, sliding friction forces, etc.) and environmental effects (e.g., gravity, viscosity, etc.); and
    \item constraint-based modeling (CBM), which uses additional geometric constraints to locate the parts in the assembly configuration by artificially reducing the DOF of the manipulated objects, similar to a CAD system.
\end{itemize}
Although such a distinction is popular in the literature with slightly variant articulations, it is imprecise and often misleading. This becomes clear by noting that motion dynamics is essentially a constrained optimization problem in disguise, hence there is no fundamental distinction between PBM and CBM as defined by the above statements. In particular, rigid body dynamics is given by the Lagrangian formulation of the equations of motion, where the requirement of no-collision between solids is equivalent to a holonomic unilateral constraint \cite{Lysenko2013}. Impulse forces and torques then originate from the Lagrange multipliers associated to these constraints by Gauss' principle of least constraint \cite{Moreau1985,Redon2002a}. Therefore, there is no fundamental difference, in terms of the underlying physics and mathematics, between finding the motion trajectories under the effect of contact forces and torques (i.e., PBM), on the one hand, and limiting the motion DOF by additional constraints (i.e., CBM) on the other hand. In fact, most implementations use both PBM and CBM, even though one of them might appear as the dominant theme.

\paragraph{Constraining Schemes.}

One can still draw a more meaningful classification based on how the constraints are formulated in practice from a knowledge of individual part geometries and their spatial relations. Based on a review of different techniques in the literature, we believe that the following provides a more precise and consistent definition with respect to the existing methodologies:
\begin{itemize}
    \item The first approach uses `physical constraints' defined as the set of constraints that arise {\it organically} from part geometries (e.g., holonomic constraints due to no-collision condition) and kinematics (e.g., nonholonomic constraints due to sliding motion specifications); whereas
    \item The second approach uses `geometric constraints' or `kinematic constraints' introduced {\it artificially} to replace the collision response and simplify the solution, ranging from manually specified `virtual fixtures' to heuristically identified `mating constraints' (e.g., co-planarity, co-axiallity, vertex/edge/face coincidence, distance or angle offsets, etc.).
\end{itemize}
Once again, we do realize that the use of the adjectives `physical/geometric/kinematic' for the classes of constraints still bears the possibility of confusion. The so-called physical constraints are directly imposed due to the interplay between geometry and kinematics of different objects, and the artificial constraints are also solved by appealing to physics-based dynamic simulation.%
\footnote{It is possible to devise purely kinetostatic constraint solvers by ignoring the dynamic effects---i.e., solving 1st-order, rather than 2nd-order differential equations. However, realistic simulation requires taking dynamic effects into account.} 
Therefore, the keywords ``organic'' versus ``artificial'' would perhaps constitute more meaningful adjectives for the different types of constraints classified according to this scheme. Nevertheless, we use the former terminology (with some extra care) for the sake of consistency with the conventions in the literature.

\begin{figure}
    \centering
    \includegraphics[width=0.48\textwidth]{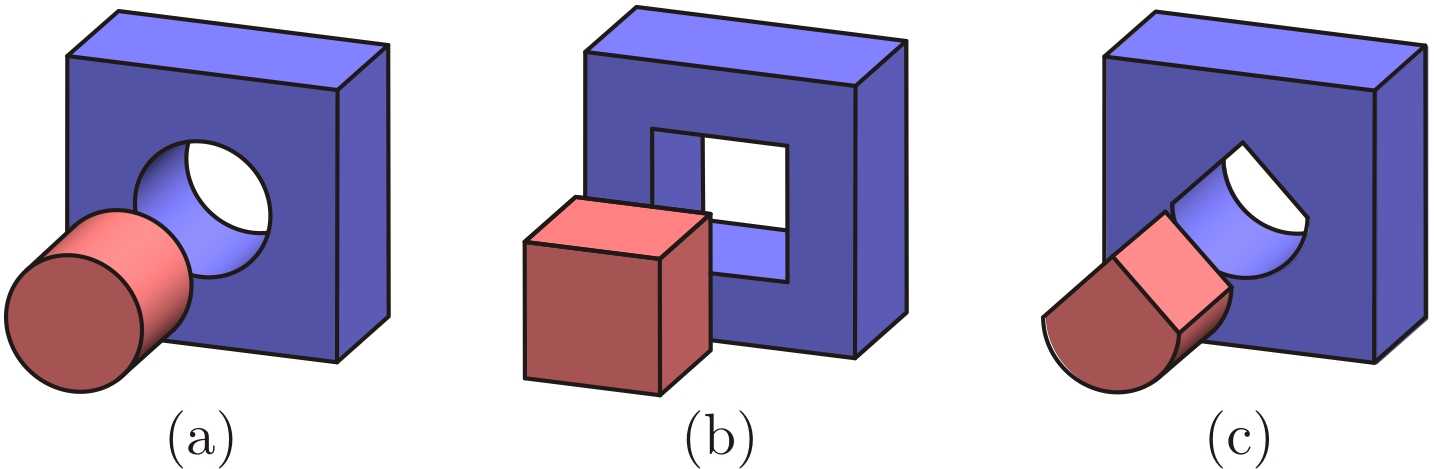}
    \caption{Examples of peg-in-hole assembly problems. Figure adopted from \cite{Behandish2014a,Behandish2015}.} \label{figure4}
\end{figure}

\begin{examp} \label{examp_peghole}
    Consider the simple peg-in-hole assembly examples shown in Fig. \ref{figure4}. One could use a variety of geometric representations including exact (e.g., parametric B-rep) or approximate (polygonal mesh or voxelization) representations to solve for the no-collision constraints. In this case, these `physical constraints' arise organically from the part geometries and are implicitly accounted for by ensuring---through the application of contact forces and torques---that the intersection volume of the two parts remains zero at all times. However, whenever the contact geometry is simple---e.g., cylindrical as in Fig. \ref{figure4} (a) or prismatic as in Figs. \ref{figure4} (b, c)---it is possible to simplify the problem by restricting the motion DOF (e.g., from the original 6 DOF to 1, 2, or 3 DOF) by artificially introducing `geometric constraints'. In this case, the complex problem of intersection test between arbitrary shapes is reduced to that of simpler geometric abstractions (i.e., virtual fixtures) such as incidence relations between axis lines and corner points. Such incidence relations can be enforced by virtual (axial and torsional) spring-damper couplings whose equilibrium states correspond to proper alignment of virtual fixtures.
\end{examp}

\paragraph{Kinematic Pairs.}

A more careful examination of the type of assembly problems similar to the above example yields a more rigorous classification of the assembly simulation methods based on the type of `kinematic pairs' \cite{Reuleaux2012}:
\begin{itemize}
    \item Lower kinematic pairs correspond to surface contact between parts and are classified completely into the well-known six classes \cite{OConnor1996}---namely, revolute, prismatic, helical, cylindrical, spherical, and planar. Each class corresponds to an automorphism (i.e., symmetry subgroup) of the 6D configuration space ($\conf-$space) of rigid motions. 
    \item Higher kinematic pairs correspond to curve or point contact between parts, for which no such classification exists. For pair of objects of arbitrary shapes, the type of contact belongs to this class in general.
\end{itemize}
For lower kinematic pairs, the aforementioned classification enables one to encode all possible interactions between a given pair of parts into well-defined types of `mating constraints' as is customary in many CAD systems. Examples are co-planarity, co-axiallity, co-centricity, and similar conditions. These constraints can be simplified into incidence relations between lower-dimensional geometric constructs that abstract the type of contact. For instance, the cylindrical pair in Fig. \ref{figure4} (a) can be abstracted by the incidence of the axis of the peg (or at least two points on it) with the axis of the hole, which constrains the motion to a 2D subgroup of $\SE{3}$; namely, a translation along and a rotation about the same axis. The enforcement of this constraint in a VE can be realized by a spring-damper coupling between the two cylindrical axes (or certain points on them) that resists an increase in the angle between the two, but is indifferent to the relative rotation around the axes. Similarly, the prismatic pairs in Fig. \ref{figure4} (b, c) can be captured by additional conditions to lock the rotation around the axes and restrict the motion further into a 1D subgroup of $\SE{3}$. See also Fig. \ref{figure7} in Section \ref{sec_geomconst} for a similar method \cite{Tching2010,Tching2010a}.

An examination of the different existing implementations reviewed in Appendix \ref{app_VA}, reveals that most PBM-based simulations use CD to identify the repulsive effects that resist the penetration of individual parts, making direct use of geometry (i.e., exact or approximate representations). On the other hand, most CBM-based simulations disregard the explicit geometric information and use the additional mating constraint semantics imported from the CAD models to implement the spring-damper couplings that contribute both attractive and repulsive effects. An exception to this theme is the \textsf{Snap-to-Fit} system by Olsson et al. \cite{Olsson2013} which makes direct use of explicit geometric information to create attractive and repulsive effects between the parts that are close to each other without penetration. More specifically, each point on the moving part's surface is coupled to the nearest neighbor on the stationary part's surface using a virtual spring-damper to create the snapping effect. Although it applies to arbitrary geometry and does not depend on simplifying assumptions on the contact features and kinematic pairs, such a simplistic `magnetic' energy model is often counterintuitive and countereffective with regard to the assembly intent. In particular, the underlying energy field merely attempts to bring the parts to proximity by pairing the closest points on their respective surfaces and is indifferent to the geometric constraints like the ones implied implicitly in the peg-in-hole examples of Fig. \ref{figure4}. In contrast, Behandish and Ilie\c{s} offered an energy model in \cite{Behandish2014a,Behandish2015,Behandish2015a,Behandish2016,Behandish2016a} that exhibits stronger relationships with both physical (i.e., collision resistant) and geometric (i.e., mating induced) constraints and blends the two in a single formulation. More details on this unified approach are given in Section \ref{sec_unified}.

\subsection{Physically-Based Modeling} \label{sec_PBM}

In order to realistically simulate the dynamic interactions between parts and subassemblies (including the user's interface object/avatar), the majority of haptic-enabled assembly systems perform an explicit real-time integration of the 2nd-order differential equations of motion. These equations are either formulated as Newton+Euler's equations---e.g., as in \cite{Baraff1995,Sauer1998,Wang1998,Wang2001,Bender2014}---or Lagrange's equations---e.g., as in \cite{Lotstedt1984,Anitescu1999,Stewart2000,Tching2008,Tching2008a,Fetecau2003,Renouf2005}---both of which are equivalent in terms of the underlying mathematics, but offer different computational procedures.

The dynamic simulation typically runs at lower rates (e.g., $100$ Hz) compared to the haptic rendering loop at the device level, and the two are interfaced using a `virtual coupling' \cite{Colgate1995,Adams1999,Adams2002,Hannaford2002,Garbaya2009,He2013}, which is essentially a spring-damper model that connects two virtual instances of an object, one residing in the physics simulator and the other assumed to be attached to the user's interaction point. 

Among the earliest attempts for using PBM in haptic assembly were earlier implementations of \textsf{VEDA} \cite{Gupta1995,Gupta1997}, \textsf{VADE} \cite{Jayaram1997,Jayaram1999,Jayaram1999a}, and \textsf{HIDRA} \cite{McDermott1999,Coutee2001,Coutee2002} systems. Examples of more recent PBM-based systems are \textsf{SHARP} \cite{Seth2005,Seth2006,Seth2008,Seth2007,Seth2010} and \textsf{HAM(M)S} \cite{Lim2006,Lim2006a,Lim2007a,Lim2010,Lim2007,Badillo2013,Badillo2014}. These systems (and a number of others) were reviewed in more detail in Appendix \ref{app_VA}. The most challenging set of computations in PBM are due to solving physical constraints arising from contact between different objects in the scene, including rigid and flexible parts and subassemblies (typically imported from complex CAD models).

\subsubsection{Physical Constraints} \label{sec_physcons}

There are two common approaches for computing the contact forces and torques enforcing the CD-induced physical constraints in real-time:
\begin{enumerate}
    \item The first method, referred to as the `penalty method', uses simple force and torque models that make {\it explicit} use of collision response---e.g., a linear spring-damper model for computing the normal contact forces proportional to a measure of penetration between objects (or their offset shells) \cite{Fisher2001,Hasegawa2003,Hasegawa2004} and a proper friction model using the normal pressures and the relative sliding/rolling kinematics to compute the tangential forces  \cite{Baraff1994,Mirtich1995,Hayward2000}.
    \item The second method, referred to as `constraint-based'---another unfortunate terminology that contributes to confusion with the CBM concepts%
        \footnote{This is a possible ground for confusion (due to the bad terminology of PBM vs. CBM) as the naming similarity suggests that the PBM class is confined to explicit penalty methods only while the implicit constraint-based methods should be classified under CBM. However, this in not the case since CBM deals completely with a different family of constraints added artificially on top of the physical constraints, as pointed out in Section \ref{sec_motion}.}%
        ---instead takes an {\it implicit} account of the unilateral contact constraints and solves the more complex set of constrained equations of motion, using non-smooth Lagrangian mechanics \cite{Fetecau2003,Renouf2005,Tching2008,Tching2008a}.
\end{enumerate}
The penalty method is easy to implement and fast to integrate---given an efficient collision response and impact/friction modeling algorithm---due to the simple form of unconstrained Newton+Euler's or Lagrange's equations of motion. However, the robustness of the penalty method is heavily dependent on small integration time-steps to ensure minimal violations of constraints and rapid response to correct them. This is difficult to achieve with accurate elasticity models for impact mechanics, inferring normal forces from penetration depth as well as friction models for rolling/sliding mechanics, inferring tangential forces from relative kinematics. The constraint-based method, on the other hand, is more difficult to implement and takes more computing time due to the solution of complicated differential equations, especially as the number of contact features increases. However, it produces more accurate and reliable results, avoids the overhead due to predicting the penetration depth for collision response, and provides straightforward means to model tangential friction forces \cite{Lotstedt1984,Stewart2000}. Both methods are dependent on collision detection (CD), although they might use different CD information such as minimum distance, intersection volume, interpenetration depth, contact normal vector, etc.

\subsubsection{Collision Detection} \label{sec_CD}

\begin{figure*}
    \centering
    \includegraphics[width=0.85\textwidth]{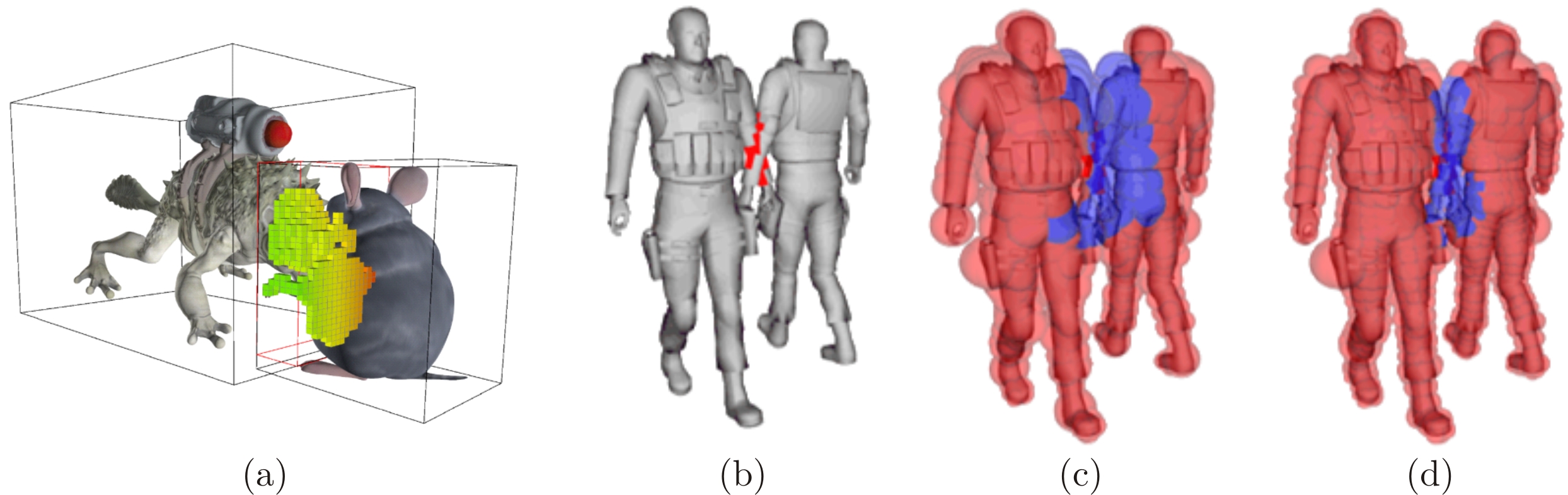}
    \caption{Combinatorial CD methods use auxiliary approximate representations ranging from voxel maps (a) to bounding spheres (b--d). Figure courtesy of Nie{\ss}ner et al. \cite{Niessner2013} and Kavan and Zara \cite{Kavan2005}.} \label{figure5}
\end{figure*}

There are several surveys of CD methods (in a general context) for rigid bodies \cite{Lin1998,Jimenez2001,Kockara2007} and flexible elements \cite{Teschner2005}.%
\footnote{For a good collection of collision detection and proximity query packages, visit the website of the GAMMA research group (Lin et al. and Manocha et al.) at the University of North Carolina at Chapel Hill: \href{http://gamma.cs.unc.edu/research/collision/}{gamma.cs.unc.edu/research/collision}.
Here is a list of references in chronological order: \textsf{I-COLLIDE} \cite{Cohen1995}, \textsf{V-COLLIDE} \cite{Hudson1997}, \textsf{RAPID} \cite{Gottschalk1996}, \textsf{V-Clip} \cite{Mirtich1998}, \textsf{IMMPACT} \cite{Wilson1999}, \textsf{H-COLLIDE} \cite{Gregory1999,Gregory2005}, \textsf{PQP} \cite{Larsen2000}, \textsf{SWIFT} \cite{Ehmann2000}, \textsf{SWIFT++} \cite{Ehmann2001}, \textsf{PIVOT} \cite{Hoff2001,Hoff2002}, \textsf{DEEP} \cite{Kim2002}, \textsf{CULLIDE} \cite{Govindaraju2003}, \textsf{DEFORMCD} \cite{Govindaraju2005}, \textsf{DVD} \cite{Sud2006}, and \textsf{SELF-CCD} \cite{Tang2010}.}
Here we restrict ourselves to a brief review of the most popular methods for real-time computations.

The classical polyhedral CD methods were used in the earliest systems for haptic assembly. Examples are Voronoi-clipping/marching methods---e.g., \textsf{V-Clip} \cite{Mirtich1998}, \textsf{SWIFT} \cite{Ehmann2000}, and \textsf{SWIFT++} \cite{Ehmann2001} used in \textsf{HIDRA} \cite{McDermott1999,Coutee2001,Coutee2002}. For realistic applications with geometric complexities that require mesh approximations with large polygon counts (i.e., in the order of millions of triangles), these methods are not fast enough to support the 1 kHz haptic rendering.

On the other hand, the bounding volume hierarchy (BVH) methods have been among the most popular CD methods for graphic and haptic rendering purposes. These methods operate by approximating virtual objects recursively with hierarchies of simple bounding shapes (offering fast collision predicates) stored in a tree-like data structure. This allows for quickly ruling out early miss configurations in the broadphase CD as well as a trade-off mechanism between accuracy and running time; namely, by proceeding deep enough down the tree to consume as much of the $\sim 1$ millisecond per haptic frame as made available for CD budget. Examples are axis-aligned bounding box (AABB) tree-based methods---e.g., a simple algorithm in \cite{Morris2006} used in an early haptic training platform \cite{Morris2007}---and oriented bounding box (OBB) tree-based methods---e.g, \textsf{I-COLLIDE} \cite{Cohen1995} and \textsf{V-COLLIDE} \cite{Hudson1997} giving birth to the \textsf{H-COLLIDE} \cite{Gregory1999,Gregory2005} haptic module, and \textsf{RAPID} \cite{Gottschalk1996} used in \textsf{MIVAS} \cite{Wan2004,Zhu2004}. Among other BVH methods are discrete oriented polytope (DOP) tree-based methods \cite{Zachmann1998,Klosowski1998} which are generalizations of AABB trees with cubic bounding boxes to convex polytope bounding volumes. A more successful approach is that of hierarchical bounding sphere (HBS) tree-based methods \cite{Hubbard1993,Quinlan1994,Palmer1995,Hubbard1995,Hubbard1996,OSullivan1999,Dingliana2000,Bradshaw2004}, which use bounding spheres (as the name indicates) instead of boxes or other polyhedra, including variants such as the bounded deformation (BD) trees \cite{James2004}. The spherical symmetry of the primitives in all levels of the HBS trees offers simple and fast radial-based collision predicates that are invariant under rotations, making them popular in more recent haptic implementations \cite{Ruffaldi2008,Chang2010}.

For a long time, uniform volumetric enumeration methods such as the one used in \textsf{Boeing Corp.}'s \textsf{Voxmap PointShell (VPS)}$^\text{TM}$ library \cite{Wan2003,McNeely2005a,McNeely2005,McNeely2006} became very popular for VR applications \cite{Johnson2001,Renz2001,Kim2003,Kim2004}. \textsf{VPS}$^\text{TM}$ works by testing the moving objects represented by a shell of vertices and normals (i.e., the `pointshell') against the stationary obstacles represented by a map of voxels (i.e., the `voxmap'), and was used in the earlier versions of \textsf{SHARP} \cite{Seth2006,Seth2008}. Several improvements were proposed to the \textsf{VPS}$^\text{TM}$ method, ranging from model enhancements---e.g., by using signed distance fields to enhance continuous force and torque response \cite{Barbic2007,Barbic2008}---to implementation speed-ups---e.g., by using improved data structures \cite{Sagardia2008,Sagardia2014}. Although still being popular due to their simplicity and efficiency, the approximate nature of discrete volumetric representations makes them ineffective for low-clearance assembly \cite{Seth2006,Seth2008,Seth2007,Seth2010}. To overcome this, later versions of \textsf{SHARP} \cite{Seth2007,Seth2010} employed the \textsf{Collision Detection Manager (CDM)} module of \textsf{Siemens}' \textsf{D-Cubed}, which makes direct use of exact B-rep information extracted from the CAD models. Of course, this comes at the expense of slowing the CD process down and making it impractical for large and complex models (e.g., with numerous NURBS patches).

Coutee and Bras \cite{Coutee2002} compared multiple polygon-based CD toolboxes---namely, \textsf{V-Clip} \cite{Mirtich1998}, \textsf{SWIFT} \cite{Ehmann2000}, and \textsf{SWIFT++} \cite{Ehmann2001}---with \textsf{VPS}$^\text{TM}$ \cite{Wan2003,McNeely2005a,McNeely2005,McNeely2006} in terms of their features and capabilities to provide closest point, collision features, penetration depth, geometric constructions, multibody detection and their effectiveness for haptic simulation. They argued that the lone advantage of \textsf{V-Clip} over the other algorithms is that it provides (a not-so-accurate measure of) penetration distance, which can be overcome by using simple tricks via \textsf{SWIFT(++)}. On the other hand, \textsf{VPS}$^\text{TM}$ and \textsf{SWIFT++} have the attractive feature of handling arbitrary nonconvex objects, while \textsf{V-Clip} and \textsf{SWIFT} can only handle nonconvex objects as collections of convex pieces. Kim and Vance \cite{Kim2003,Kim2004} conducted a more inclusive study comparing a larger number of CD packages---namely, \textsf{I-COLLIDE} \cite{Cohen1995}, \textsf{V-COLLIDE} \cite{Hudson1997}, \textsf{RAPID} \cite{Gottschalk1996}, \textsf{PQP} \cite{Larsen2000}, and \textsf{SOLID} \cite{Bergen1997} in addition to the aforementioned four---in terms of their query types and response times. The study concluded that \textsf{VPS}$^\text{TM}$ is a better choice due to its ease of CAD preprocessing, faster CD query response, and ability to model physical interactions between parts for force feedback---hence its broad popularity ever since for developing haptic rendering systems.

A promising method was recently developed based on hierarchical inner sphere tree (IST) packing \cite{Weller2009,Weller2011,Weller2013} and successfully applied to PBM for haptic rendering \cite{Weller2009a,Weller2009b}. In contrast to the bounding sphere methods described above, the IST packing algorithms use the hierarchy of spheres to pack the {\it interior} of the virtual objects---altering the topological structure. Nevertheless, both methods share the computational advantage due to the spherical symmetry in primitive collision predicates. The sphere-packing approximation of the interior can be viewed as a nonuniform extension to the uniform volumetric enumeration approach used in \textsf{VPS}$^\text{TM}$, since it starts from a grid-based discretization of the shape (similar to the voxmap in \textsf{VPS}$^\text{TM}$) over which the distance function is computed and the sphere centers are populated using a simple greedy algorithm. It was shown to outperform the \textsf{VPS}$^\text{TM}$ for nonconvex moving objects, but its effectiveness to handle thin objects is yet to be tested \cite{Perret2013}.

\paragraph{Discrete vs. Continuous Methods.}

In contrast to the aforementioned {\it discrete} methods that evaluate the collision certificate at intermittent integration time-steps, others have developed {\it continuous} methods based on OBB trees \cite{Redon2002,Redon2004,Redon2005,Zhang2006,Zhang2007,Zhang2007a}, which attempt to interpolate the first instance of contact in between two subsequent time points along the dynamic time-stepping. This enables speeding up the integration of constrained motion by decreasing the number of unilateral constraints using a fast `clash detection' algorithm based on relative motion during each time-step \cite{Perret2013}. Although this method was used in an operational context in the industry, it was soon abandoned due to a lack of commercial support \cite{Perret2013}.

\begin{figure*}
    \centering
    \includegraphics[width=0.65\textwidth]{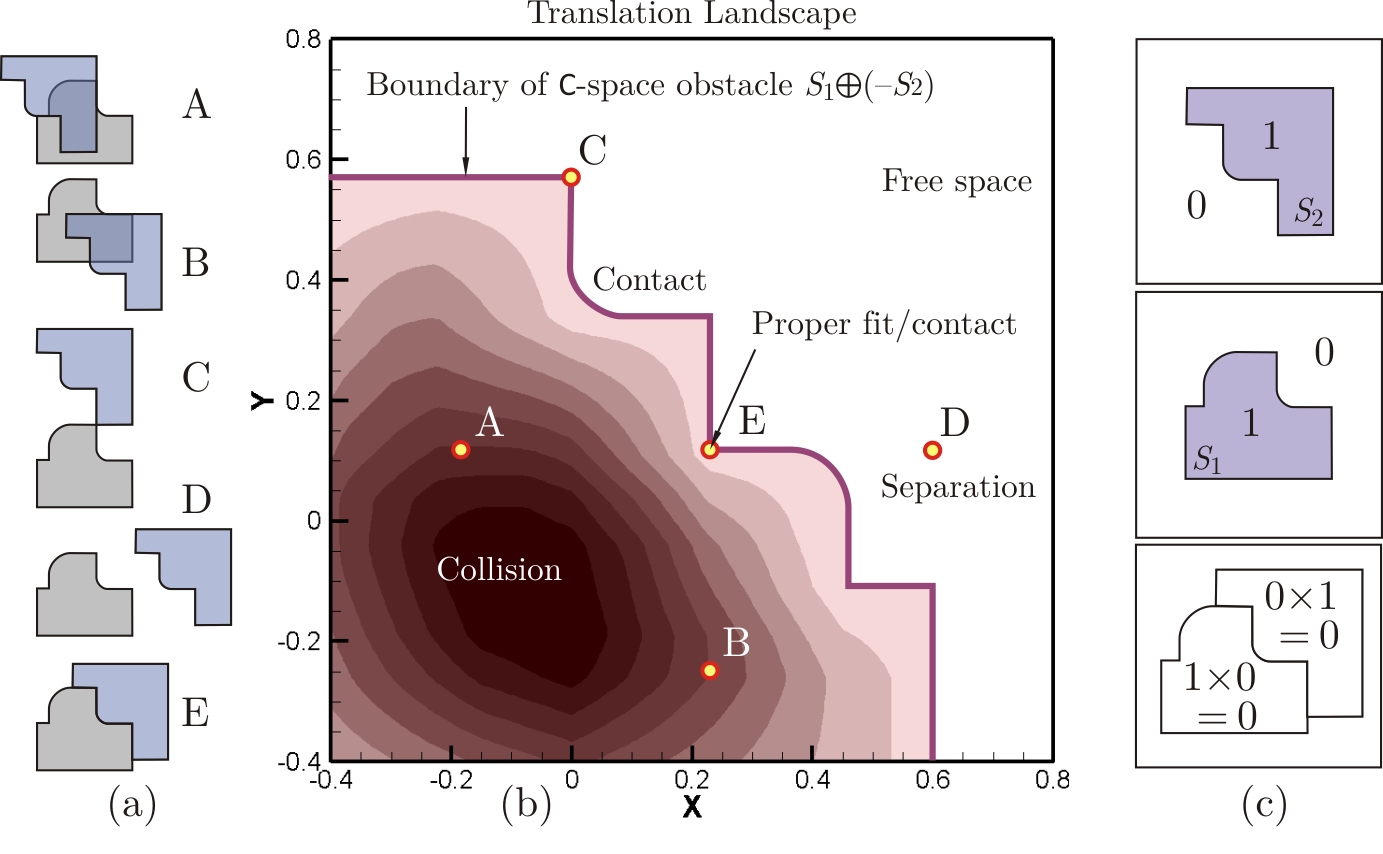}
    \caption{Different configurations (a) are evaluated using a gap function (b), based on overlapping indicator functions (c), formulated as a cross-correlation. Compare to Fig. \ref{figure9}.} \label{figure6}
\end{figure*}

\paragraph{Combinatorial vs. Analytic Methods.}

At a very abstract level, one could classify CD methods into combinatorial techniques (including most, if not all of the aforementioned methods), and analytic methods \cite{Lysenko2013,Behandish2015d}. PBM applications require not only a collision/non-collision certificate, but a gradient of the constraint function to compute the contact forces and torques. A disadvantage of the combinatorial methods is their indirect approach to infer a gradient-like quantity from the certificate point, which is not trivial for general surface contact \cite{Lysenko2013}. The analytic methods \cite{Comba1968,Gilbert1985,Moreau1985}, on the other hand, provide a more uniform and robust alternative, which has been popular for a long time in robotics \cite{Lozano-Perez1983}.

Recently, Lysenko \cite{Lysenko2013} developed an efficient analytic method for CD based on earlier works in robotics spatial planning \cite{Kavraki1995,Curto2002} and group morphology \cite{Roerdink2000,Lysenko2010,Lysenko2011a}. The method formulates the so-called `gap function' \cite{Harker1990} which measures the penetration volume as a {\it convolution} of nonnegative real-valued defining functions---e.g., the indicator (i.e., characteristic) functions, bump functions, etc.---of the two objects over the configuration space ($\conf-$space) of their relative motion. By taking advantage of the powerful {\it convolution theorem}, the method applies Fourier transforms to compute the collision response and its general $\conf-$space gradient (both with respect to translations and rotations) for narrowphase CD. After sampling the defining functions enumerated over a uniform grid (i.e., voxel map), the computations can be carried out rapidly in parallel (on CPU or GPU) using the radix-2 fast Fourier transform (FFT) \cite{Cooley1965}.

More recently, Behandish and Ilie\c{s} \cite{Behandish2015d} extended this method to take advantage of the more time- and memory-efficient spherical decompositions---as in HBS and IST methods described earlier. The idea is to replace the grid-based uniform sampling of the defining functions with nonuniform grid-free samplingover a grouping of 3D balls of different sizes, viewed as a 3D slice of a grouping of 4D cones of the same size, which, in turn, can benefit from the convolution paradigm and the nonequispaced fast Fourier transforms (NFFT) \cite{Potts2001}. This follows some of the most effective computational methods in state-of-the-art protein docking \cite{Bajaj2011,Bajaj2013} and enables combining the strengths of combinatorial sphere-trees and analytic convolutions.%
\footnote{The drawback is that unlike classical FFTs that are implemented as efficient, stable, easy-to-use and widely available open-source packages on the CPU (e.g., \textsf{FFTW} \cite{Frigo1998,Frigo2005}) and the GPU (e.g., \textsf{cuFFT(W)}), similar implementations for NFFTs are scarce \cite{Keiner2009,Kunis2012}.}

The analytic methods have great potential for haptic applications, as they are indifferent to the topological, geometric, and syntactic complexities of the colliding objects, and provide a graceful trade-off mechanism between fidelity and running time by `low-pass filtering' and `anti-aliasing' of the Fourier transforms of the individual objects \cite{Lysenko2013,Behandish2015a,Behandish2016}. The latter is particularly appealing for haptic application with a sub-millisecond budget for CD and rending altogether, which dictates an upperbound on the number of dominant modes in the frequency domain representations of shape descriptors. Although at very high accuracies the FFT-based CD can become slower than other exact techniques, the CD cost scales with desired accuracy in a predictable and well-define manner, irrespective of geometric complexity. Thus one can fine-tune and calibrate the low-pass filter by specifying the available time budget, querying the hardware capacity, and deciding on the accuracy trade-off without worrying about the actual geometric details. As more efficient high-performance hardware solutions become available, the filter window function can be enlarged systematically to achieve higher accuracies without violating the frame rate requirements.

In principle, one could use the analytic methods based on convolution and Fourier filtering for time-critical narrowphase CD instead of (or in combination with) traditional combinatorial CD methods for guiding haptic assembly and disassembly. However, we are not aware of their implementation into any software library or simulation engine at the time of writing this article. Nevertheless, the difficulties faced in low-clearance assembly will continue to exist as a natural result of input data noise due to device inaccuracies and hand vibration, leading to unstable dynamic response and undesirable `buzzing' in haptic feedback.

\begin{examp} \label{examp_Cspace}
    Figure \ref{figure6} (b) illustrates the translational $\conf-$space landscape for a pair of 2D solids shown in panel (a), along with the colormap for the gap function measuring the intersection area. To make the illustration possible, the motion is restricted to translation only, i.e., the landscape in panel (b) is a section (corresponding to zero rotation) through the full 3D $\conf-$space. Each of the relative positions in panel (a) are represented by a point in panel (b). The gap function $f : \RRR \to [0, +\infty)$ formulated as the convolution of part indicator (i.e., characteristic) functions penalizes collision as in positions A and B (i.e., $f(\bx) > 0$), but does not differentiate point contact in C and separation in D from proper fit/contact in E (i.e., $f(\bx) = 0$). This is clearly due to the fact that the convolution integral is only able to measure properties that correspond to full-dimensional (i.e., Lebesgue-measurable) intersections while those of lower-dimensional contact regions vanish during the course of integration.%
    \footnote{This is a major caveat to keep in mind when using analytic approaches in general which only allow a one-to-one correspondence between regularized morphological concepts and measure-theoretic equivalents \cite{Lysenko2010,Lysenko2011a}. A solution paradigm by introducing Dirac delta calculus into the analytic methods is given in depth in \cite{Behandish2016b}.}
\end{examp}

The gap function approach to CD by convolving characteristic functions \cite{Lysenko2010,Lysenko2011a,Lysenko2013} was extended to formulate a generic `geometric energy' field \cite{Behandish2016a} by convolving skeletal density functions (SDF) \cite{Behandish2014a,Behandish2015}, which, in addition to accounting for the {\it repulsive} effects and resist inter-penetrations, incorporates {\it attractive} effects that take over when there is no collision and favor proper fit/contact configurations exhibiting better local {\it shape complementarity} \cite{Behandish2014} (Fig. \ref{figure8}). The method can still benefit from CPU- and GPU-accelerated FFT computation and on-the-fly filtering \cite{Behandish2015a,Behandish2016} for sub-millisecond rendering budget in 1 kHz haptic assembly. More details on this method are given in Section \ref{sec_unified}.

\paragraph{Rigid vs. Flexible.}

A discussion of real-time CD methods would not be complete without a reference to contact modeling between deformable objects, which are prevalent in industrial assembly (e.g., electric cables, hydraulic hoses, rubber seals, leather furnishings, etc.) \cite{Perret2013}. Classical methods range from discrete mechanical elements (DME) \cite{Bridson2002,Bridson2002a,Fuhrmann2003} to finite element method (FEM) \cite{Zhuang1999,Hirota2000,Irving2004,Picinbono2003}, which are suboptimal for high frame rate haptic rendering. A few recent studies have successfully applied Signorini's contact model to haptic assembly of deformable objects \cite{Duriez2004,Duriez2004a,Duriez2006}. Nevertheless, handling complex deformable shapes and large models remains a challenge \cite{Perret2013}.

\paragraph{Physics Simulation Engines (PSE).}

Combinations of the aforementioned combinatorial algorithms have been implemented into popular physics simulation engines (PSE) for graphic and haptic rendering, such as \textsf{Ageia PhysX}$^\text{TM}$ SDK \cite{Chan2009} and \textsf{Bullet Physics} SDK \cite{Sagardia2014}. Gonzalez-Badillo et al. \cite{Badillo2014a,Badillo2014b} recently conducted a comparative performance evaluation of these PSEs and their CD capabilities in practice for haptic assembly. In particular, they measured and compared the task completion time (TCT), mean force feedback (MFF), and physics simulation time (PST) indices in several benchmark examples for the static \textsf{trimesh/HACD} module of \textsf{PhysX}$^\text{TM}$ versus the \textsf{GIMPACT} module of \textsf{Bullet} and concluded that in general the latter outperforms the former for haptic assembly tasks. An important exercise would be to
\begin{enumerate}
    \item develop robust implementations of analytic methods into the existing PSEs or stand-alone open-source and widely-available libraries; and
    \item conduct similar comparative studies to understand their strengths and limitations, and the opportunities to blend them together with combinatorial techniques.
\end{enumerate}

\subsection{Constraint-Based Modeling} \label{sec_CBM}

\begin{figure*}
    \centering
    \includegraphics[width=0.85\textwidth]{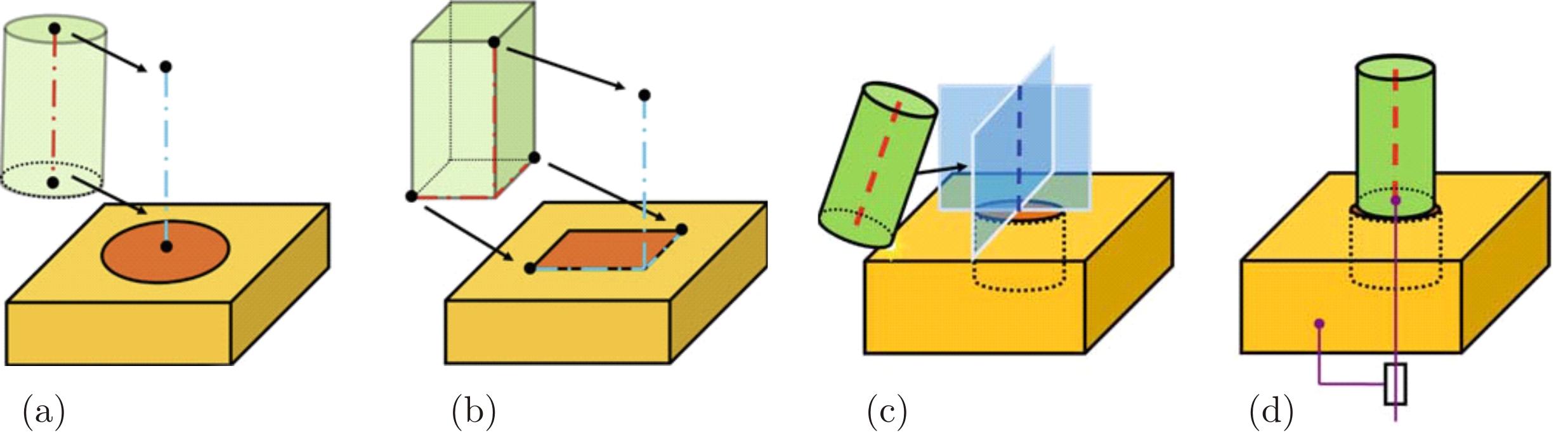}
    \caption{Using virtual fixtures for abstracting the mating constraints for cylindrical and prismatic pairs (a, b). At the close vicinity of the insertion site, the virtual fixtures (e.g., guiding planes in (c)) are used to align the peg along the hole. Once the proper alignment is reached, the mating pair's DOF is limited to model a suitable mechanical joint (e.g., cylindrical pair in (d)). Figure courtesy of Tching et al. \cite{Tching2010,Tching2010a}.} \label{figure7}
\end{figure*}

Although implementing physics-based simulation with a combination of CD and impact/friction mechanics seems the most natural choice (at least in theory) for a virtual mimicry of real-world constrained motion, it is not reliable in practice for final insertion of the objects into position \cite{Seth2006,Seth2008,Seth2007,Seth2010,Vance2011,Perret2013}.

At a fundamental level, this happens due to the degeneracy of the collision-free feasible subspace (i.e., the `free space') in the neighborhood of the final assembly configuration, leading to decreased DOF associated with common mating constraints. Most mating constraints used to model zero-clearance mechanical joints are characterized with multiple compatible unilateral (i.e., inequality) constraints that are critically satisfied during contact leading to one or more bilateral (i.e., equality) constraints, restricting the motion to a surface or a curve.%
\footnote{An inequality constraint $g(\bx) \geq 0 ~(\bx \in \mathds{R}^3)$ is `critically satisfied' if $g(\bx) = 0$. Two constraints $g_1(\bx) \geq 0$ and $g_2(\bx) \geq 0$ are compatible if they define a nonempty subset of the $3-$space. If that subset forms a lower dimensional subspace over the intersected boundaries, then they collectively define a bilateral equality constraint $g_1(\bx) = g_2(\bx) = 0$.}
Such a degenerate collection of critically satisfied inequality constraints is extremely unstable with respect to small perturbations, rendering lower kinematic pairs and their constraint resolution practically `incomputable'. This is because a small error in constraint specification may lead to major topological changes to the collision-free configuration subspace (i.e., the `free space') that either completely eliminate the constraint or make it theoretically unresolvable unless tolerances are explicitly incorporated.

\begin{examp}
    Take, for example, two inequality constraints $g_1(\bx) \geq 0$ and $g_2(\bx) \geq 0$ corresponding to non-penetration condition for two planar surfaces between which a planar part is sandwiched without a clearance. %
    The intersection of the two constraints $g(\bx) := g_1(\bx) g_2(\bx) \geq 0$---assuming nonnegative analytic functions $g_1, g_2, g: \RRR \to [0, +\infty)$---is a 3D subspace of the 6D configurations space defining a lower kinematic pair with two translational and one rotational DOF. Upon introducing a small perturbation $\epsilon \geq 0$ (e.g., $g_{1,2}' (\bx) := g_{1,2}(\bx) \pm \epsilon$) that 3D subspace may widen up to a 6D region (e.g., in `clearance fits') or completely disappear (e.g., in `shrink fits').
\end{examp}

At a practical level, it is difficult to stabilize the motion along the degenerate subspace for at least two reasons:
\begin{enumerate}
    \item numerical geometric errors due to the approximate representations used in fast CD methods popular for haptic rendering (e.g., voxelized interior or triangulated boundary); and
    \item input noise due to authentic hand vibrations and device encoder reading errors.
\end{enumerate}
The latter is particularly important, implying that even with the improved accuracy of the algorithms that use fine mesh approximations or exact B-rep data, using CD alone for low-clearance insertion is still impractical \cite{Seth2007,Seth2010,Vance2011}. One can alleviate the problem by using smaller integration time-steps to keep the unnecessary minor collision events (i.e., violations of the equality constraints) at a minimum, to achieve more stable dynamic response and haptic feedback. However, the finite time-step in real-time applications is lowerbounded by the computation time per frame, which is dictated by hardware capacity.

An alternative solution is to {\it artificially} introduce a set of bilateral constraints, rather than relying solely on the group of unilateral constraints organically resulted from CD. Such an approach is mathematically equivalent to a local reparameterization of the configuration space to embed the feasible subspace and to ensure the fulfillment of the constraints explicitly rather than attempting to solve for the original constraints specified implicitly. These artificial constraints are typically classified into geometric (i.e., holonomic) and kinematic (i.e., nonholonomic) constraints.

\subsubsection{Geometric Constraints} \label{sec_geomconst}

In most commercially available design and assembly environments such as the modern CAD software (e.g., \textsf{CATIA}$^\circledR$, \textsf{Pro/ENGINEER}$^\circledR$, \textsf{NX}, etc.) the so-called `mating constraints' are classified into simple spatial relationships between the contact features, such as co-planarity in prismatic features, co-axiallity in cylindrical features, distance and angle offsets, etc. To solve the final insertion problem for haptic assembly, one practical approach is to {\it manually} specify the mating constraints in close proximity of the final assembly configuration. The geometric constraints can be either extracted and imported from the CAD model---i.e., manually specified by the CAD user (e.g., the designer)---or specified on-the-fly within the VE---i.e., manually specified by the VE operator (e.g., the inspector)---using a variety of constraint management systems developed for VR-CAD applications \cite{Marcelino2003,Murray2003,Murray2004}. For example, \textsf{VADE} \cite{Jayaram1997,Jayaram1999,Jayaram1999a} and \textsf{MIVAS} \cite{Wan2004,Zhu2004} directly imported pre-defined constraint information from \textsf{Pro/ENGINEER}$^\circledR$ CAD models. The later versions of \textsf{SHARP} \cite{Seth2007,Seth2010} used the \textsf{Dimensional Constraint Manager (DCM)} module of \textsf{Siemens}' \textsf{D-Cubed} for defining and solving geometric constraints within the VE itself. Rather than using the assembly semantics of the original CAD models, the virtual constraint guidance (VCG) method presented by Tching et al. \cite{Tching2010,Tching2010a} relied on user-specified `virtual fixtures' \cite{Rosenberg1993,Rosenberg1994}, which are added abstract and simple geometric elements rigidly attached to the fixed and moving parts---e.g., a pair of perpendicular planes intersecting at the axis of a cylindrical hole, to constrain and guide two points selected along the axis of a cylindrical peg. Figure \ref{figure7} illustrates the use of virtual fixtures for peg-in-hole examples with cylindrical and prismatic mating pairs.

A few recent studies attempted to {\it automatically} identify the assembly intent and associated geometric constraints by analyzing semantic information of individual part geometries \cite{Marcelino2003,Iacob2008,Iacob2011,Boussuge2012}, referred to by Vance and Dumont \cite{Vance2011} as the automatic geometric constraints (AGC) method. This method relies on matching `functional surfaces' \cite{Iacob2011}---e.g, a cylindrical surface characterized by its axis and diameter, which could be used to predict the intended mating relation and associated trajectories when a peg is brought to the proximity of a hole. However, these methods are limited to matching simple (e.g., planar, cylindrical, spherical, and conical) geometric features. The effectiveness of both VCG and AGC methods relies heavily on either manual specification of the type of mating selected from a finite library of simple constraints, or heuristic models for identifying such mating pairs when the corresponding simple geometric primitives are in proximity. A generic solution that automates the identification and pairing for features of arbitrarily complex surface geometry is missing.

\subsubsection{Kinematic Constraints}

In a similar fashion to holonomic (i.e., geometric) constraints, one could specify nonholonomic (i.e., kinematic) constraints that depend on the relative linear and angular velocities of parts during assembly. An important caveat is related to proper differentiation of the motion of the haptic proxy from discrete encoder readings in the presence of the added noise and device errors. Our experiments have led to the observation that the data filtering provided by commercial haptic device libraries and APIs might not be adequate for 1st and 2nd differentiation, and additional techniques might be required \cite{Janabi-Sharifi2000}.

\subsection{Future Directions} \label{sec_future}

Having outlined current practice and existing methodologies in haptic-assisted VP/VA (including PBM and CBM), this section briefly comments on promising directions for future studies and how the approach presented in this thesis contributes to these advancements.\footnote{Most of Sections \ref{sec_readiness} and \ref{sec_hybrid} are adopted from \cite{Perret2013,Vance2011}.}

\subsubsection{Technology Readiness} \label{sec_readiness}

Perret et al. \cite{Perret2013} presented an assessment of the maturity of the technical solutions in haptic-assisted assembly using the technology readiness level (TRL) index originally developed by \textsf{NASA} in the 1980s for space-flight systems and later adopted and expanded by the \textsf{US Air Force} to encompass other technologies. Based on this measure, they assessed the maturity of interactive rigid-body assembly simulation with haptic feedback quoted here as follows:
\begin{itemize}
    \item In terms of `path finding': TRL = 8, i.e., ``technology has proven to work in its final form and under expected conditions.''
    \item In terms of `final insertion': TRL = 5, i.e., ``the basic technological components are integrated with reasonably realistic supporting elements.''
    \item In terms of `human positioning': TRL = 6, i.e., ``representative model or prototype system, which is well beyond TRL = 5, is tested in a relevant environment.''
\end{itemize}
The `path finding' problem---more commonly referred to as `path planning' in the robotics literature \cite{Lozano-Perez1983}---is of assessing the feasibility of assembly by identifying a collision-free path (or lack thereof) from some initial configuration of a given part or subassembly to the final configuration. If no such path exists, the goal is to identify the bottlenecks and critical collision points to modify the product or process in order to resolve the design problem. The most popular automatic approach to this problem is by using probabilistic roadmap (PRM) planners \cite{Kavraki1996,Kavraki1998,Boor1999,Geraerts2004} which randomly sample the configuration space of rigid motions $\SE{3}$ and connect the nearby points to form a graph over which a feasible path can be traced. Although automatic path planning performed in an offline preprocessing step can be used later for haptic guidance as in \cite{Christiand2011,Christiand2008,Christiand2009,Hassan2010,Hassan2011,Hassan2011a,Ladeveze2010,Ladeveze2009}, the true benefit of haptic-assisted VP/VA is realized when it is used as an alternative to automatic path planning by calling upon the human's cognitive abilities and understanding of spatial relationships---e.g., to feel bottlenecks, evaluate clearances, and explore possible improvements---rather than exploiting the computational power of the machine \cite{Perret2013}. Hence it appears that basic CD-based PBM remains to be the most natural toolbox for this purpose. The TRL of 8 indicates an adequate level of maturity in this aspect, even though consistent computational improvements are needed to enable fast CD for increasing size and complexity of models. It has been shown that analytic methods similar to the one presented in this thesis are a promising future direction for both path planning \cite{Kavraki1995,Curto2002} and CD \cite{Lysenko2013,Behandish2015d}.

The `final insertion' problem appears as the least mature component with a TRL of 5, facing multiple challenges to reach effective industrial implementation. It is impossible to rely on CD alone for the final insertion of a part when low-clearance fits are involved \cite{Perret2013}. One promising future direction to solve the final insertion problem is to use a hybrid PBM+CBM approach (detailed in Section \ref{sec_hybrid}) where the traditional simulation of contact-constrained motion is confined to a `free motion' phase while fixed mating constraints are artificially added to the PSE solver for a `fine insertion' phase. Once again, these constraints are extracted from CAD model as in \cite{Jayaram1997,Jayaram1999,Jayaram1999a,Wan2004,Zhu2004}, defined by the user within the VE as in \cite{Seth2007,Seth2010,Tching2010,Tching2010a}, or determined (i.e., guessed) automatically by the system from an inspection of the geometry as in \cite{Marcelino2003,Iacob2008,Iacob2011,Boussuge2012}. One challenge faced by the hybrid approach is developing an algorithm to account for the switch between the two phases, using a variety of remedies such as blending algorithms as in \cite{Seth2007,Seth2010,Picon2008} or guiding mechanisms as in \cite{Tching2010,Tching2010a}. The second challenge is that switching off CD altogether during final insertion might lead to missing contact or collision outside the insertion site \cite{Perret2013}, to which no plausible solution has been published so far to the best of our knowledge. The alternative analytic approach \cite{Behandish2016a} unifies the two phases (detailed in Section \ref{sec_unified}) without having to deal with the challenges presented by such an artificial duality. However, in spite of its theoretical elegance and computational advantages, the latter has not been yet tested for usage in an industrial capacity.

The `human positioning' problem is largely beyond the scope of this article. One of the greatest challenges in this area is of the computational intensity of introducing an avatar into the simulation environment whose realistic model involves hundreds of new rigid bodies to deal with. See \cite{Perret2013} for more details.

\subsubsection{A Hybrid Approach} \label{sec_hybrid}

\begin{figure}
    \centering
    \includegraphics[width=0.48\textwidth]{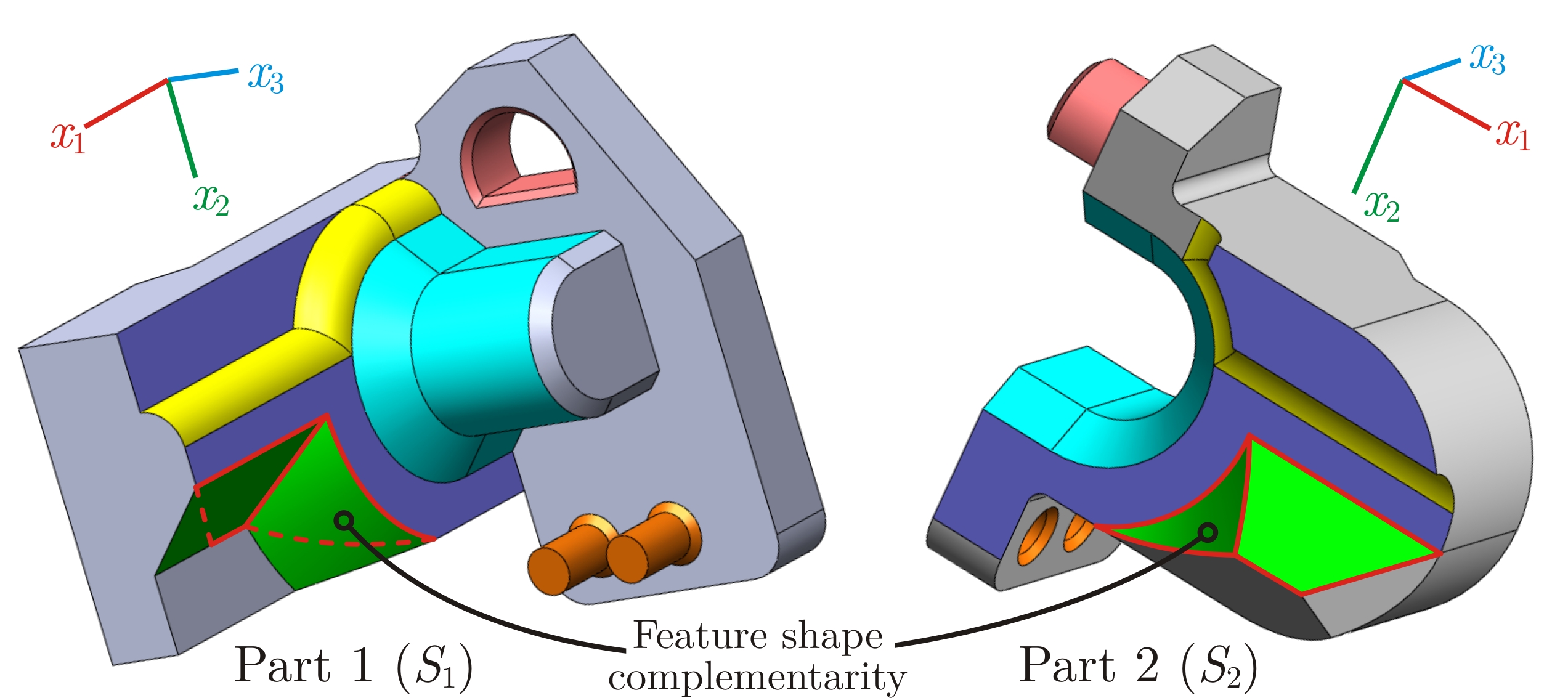}
    \caption{Even the artifacts bounded by the simplest lower-order algebraic (e.g., planar, cylindrical, etc.) surfaces can exhibit complex shape complementarity relations that are hard to detect by combinatorial matching of assembly features. Figure adopted from \cite{Behandish2015a,Behandish2016}.} \label{figure8}
\end{figure}

Although it has been shown that reducing the DOF of motion using geometric constraints supports highly accurate manipulation and positioning in VEs \cite{Bowman2001}, the ad hoc nature of the constraint detection algorithms does not provide sufficient generality to completely replace CD to constrain the motion. Consequently, the state-of-the-art in haptic assembly is a `two-phase' approach \cite{Vance2011}, i.e., to divide the process into a `free motion' phase accomplished with the help of CD engines, and a `fine insertion' phase using pre-specified or computer-predicted constraints. Vance and Dumont \cite{Vance2011} observed such a common theme among a few recent studies and systems:
\begin{enumerate}
    \item The automatic geometric constraints (AGC) method by Seth et al. \cite{Seth2007,Seth2010} relied on CD between B-rep surfaces during a free motion phase when the parts move freely (i.e., with 6 DOF except for the collision-induced constraints) in the VE. Contact between B-rep elements signaled a switch to the insertion phase of the assembly simulation when the geometric constraints were used to guide the assembly process. These geometric constraints were easily defined based on the B-rep semantics and automatically identified using simple heuristics---e.g., alignment between the cylindrical axes. Once a constraint was identified, the two parts were aligned and the number of DOF allowed for the user motion was reduced to impose the geometric constraint \cite{Vance2011}.
    \item The virtual constraint guidance (VCG) method by Tching et al. \cite{Tching2010,Tching2010a} used non-smooth dynamic simulation during an exploration phase when the parts move freely (i.e., with 6 DOF except for the collision-induced constraints) in the VE. The method relied on virtual fixtures \cite{Rosenberg1993,Rosenberg1994} to guide the moving object to a specific configuration without making any changes to the underlying CAD geometry. Once the proper insertion alignment was reached, the CD was disabled and the assembly phase was started by modeling the final insertion as a DOF-limited relative motion of simple mechanical joints (e.g., prismatic, ball, hinge, etc.). The transition between the two phases was triggered by the collision of the virtual guides between the moving and stationary objects \cite{Vance2011}.
    \item The dynamic decomposition and integration of DOF (DIOD) method by Veit et al. \cite{Veit2010} divided the assembly task into a ballistic phase and a control phase, and took a different approach based on detecting velocity changes to trigger the switch between the two phases. During the ballistic phase the user could freely move and manipulate an object at a 1:1 ratio between the tracked (i.e., device) and virtual (i.e., object) velocities. During the fine positioning phase, instead of scaling the resultant velocity to increase the dexterity, the total velocity was decomposed along 3 orthogonal directions and the scaling was applied only to the component of the velocity that is below a given threshold \cite{Vance2011}.
\end{enumerate}
There are two major difficulties faced in this approach. First, it requires developing mechanisms to detect the insertion intent (e.g., using pose or speed clues) and to model the transition between the two phases. The aforementioned implementations typically rely on CD between surface elements associated with insertion constraints \cite{Seth2007,Seth2010}, CD between the user-defined virtual fixtures \cite{Tching2010,Tching2010a}, or velocity changes that hint on the user's intent to perform an insertion task \cite{Veit2010}. Once the alignment has been reached, part CD is switched off and the number of DOF is reduced to assist the user with final insertion. Second, switching off part CD altogether is not satisfactory as a contact with geometry outside of the insertion area could oppose the movement \cite{Perret2013}. To the best of our knowledge, the latter problem is also open.

\subsubsection{A Unified Approach} \label{sec_unified}

\begin{figure*}
    \centering
    \includegraphics[width=0.65\textwidth]{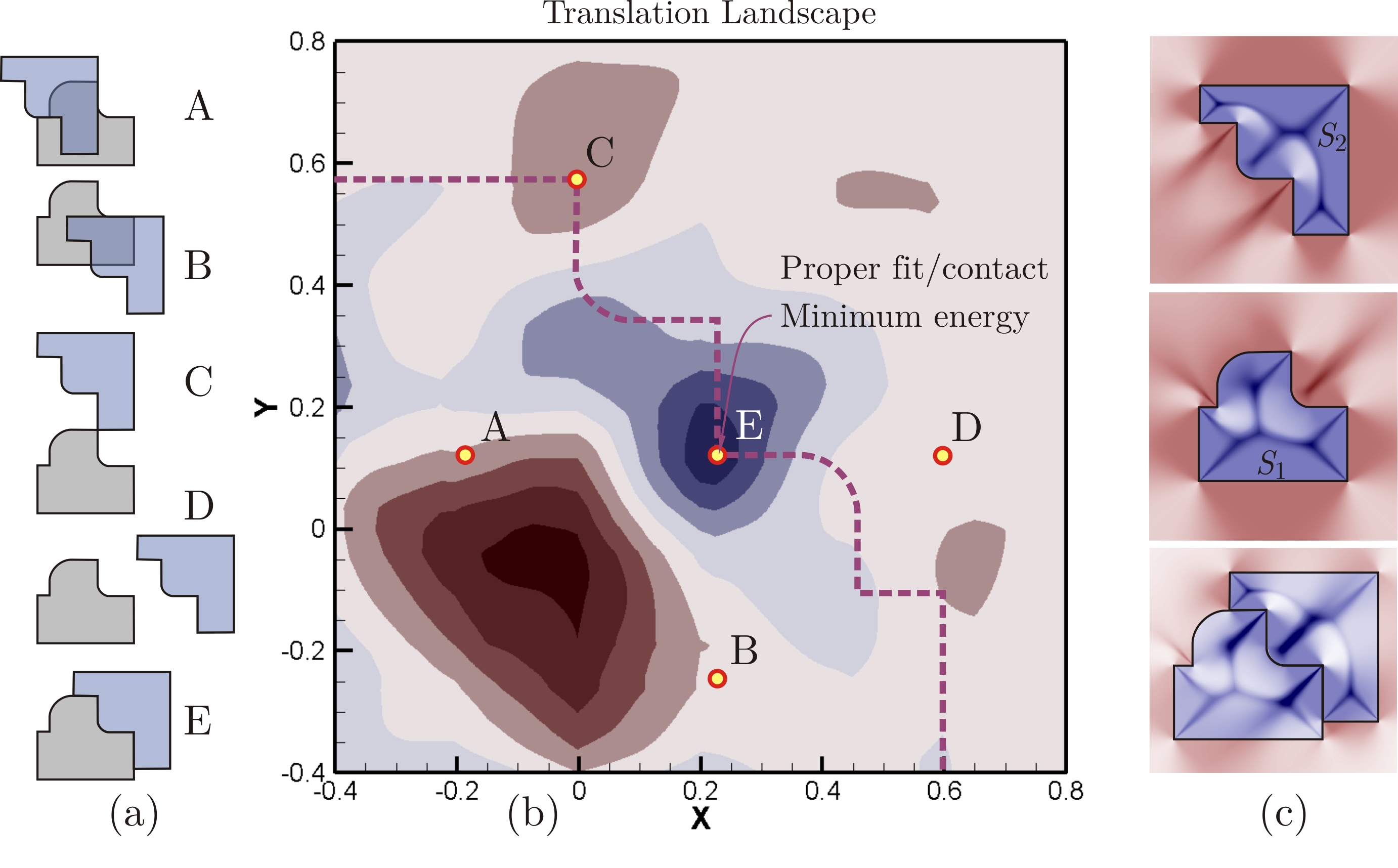}
    \caption{Different configurations (a) are evaluated using a score function (b), based on overlapping skeletal densities (c), formulated as a cross-correlation. Compare to Fig. \ref{figure6}.} \label{figure9}
\end{figure*}

Both physical constraints (i.e., captured by CD and used in PBM) and mating (geometric and/or kinematic) constraints (i.e., interpreted from CAD and used in CBM), which govern the part behavior in the two phases of the hybrid approach described in Section \ref{sec_hybrid}, are solutions to the same basic problem that lie on the two extreme ends of an spectrum in conceptual and computational terms:
\begin{itemize}
    \item The former methods apply to arbitrary geometry (i.e., general solids with semianalytic surfaces) while the latter intrinsically work only for simple shapes (e.g., assembly features with 1st- or 2nd-order algebraic surfaces).
    \item The former methods are numerically complex and time-consuming even using approximate representations (e.g., voxmap enumeration, polyhedral meshing, or BVH tree-based sampling) while the latter are as simple as computing basic and fast geometric predicates.
    \item The former methods provide a sense of crisp contact (e.g., impact and friction) at the expense of instability to small perturbations while the latter create a (sometimes unrealistically) more flexible sensation with the additional benefit of robustness to errors and noise.
\end{itemize}
Is it possible to come up with a single part behavior model that subsumes both of the above and intrinsically enables a smooth transition between them rather than salvaging one to blend the discrete phases? 

\begin{figure*}
    \centering
    \includegraphics[width=0.75\textwidth]{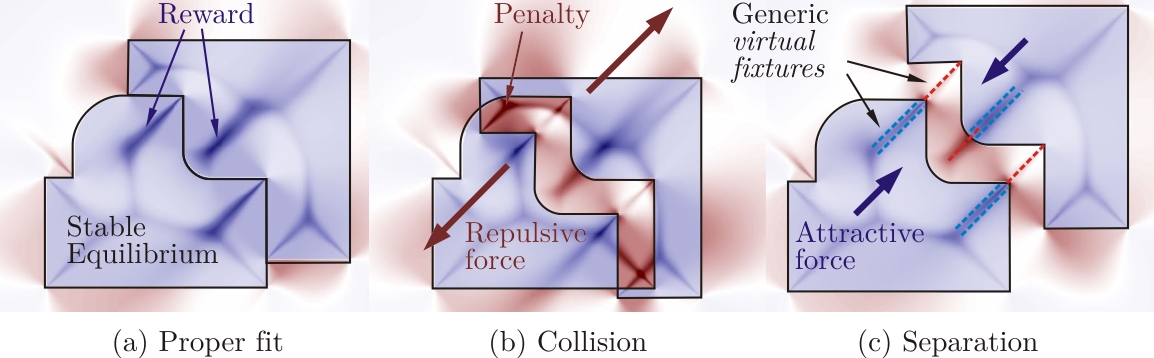}
    \caption{The interactions between SDF shape descriptors can be viewed as a generalization of virtual fixtures.} \label{figure10}
\end{figure*}

Behandish and Ilie\c{s} proposed in \cite{Behandish2014a,Behandish2015,Behandish2015a,Behandish2016,Behandish2016a} a generic and unified energy model for real-time assembly guidance that applies to objects of arbitrary shape. The formulation starts from the part geometries and {\it directly} computes the guidance forces and torques from shape descriptors of interacting features. No simplifying assumption is made on the geometry of the mating features. It was shown that implicit generalizations of the so-called virtual fixtures \cite{Rosenberg1993,Rosenberg1994} that are widely popular for haptic assembly \cite{Tching2010,Tching2010a,Vance2011,Perret2013} {\it automatically} appear in the form of interactions between skeletal density function (SDF) (Fig. \ref{figure10}). The SDF shape descriptors are piecewise continuous functions defined over the 3D space for each individual part, whose distributions capture the topological and geometric properties of the surface features that partake in assembly. The spatial overlapping of individual part SDFs---interpreted in the analytic formulation as a {\it convolution}---generates an artificial potential energy (called the `geometric energy') field \cite{Behandish2016a} which creates {\it attraction} forces and torques towards the proper alignment of assembly features. The same energy model also provides {\it repulsion} forces and torques as a natural byproduct, in the case of collisions. Therefore, it unifies the two phases of free motion and precise insertion into a single interaction mode, thus avoids the duality and switch altogether. The method subsumes analytic collision detection (CD) \cite{Lysenko2013}, and provides a generalization to analytic feature matching for geometric guidance.

\begin{examp}
    Figure \ref{figure9} repeats the gap function computation over the $\conf-$space in Example \ref{examp_Cspace} (Fig. \ref{figure6}), this time using more sophisticated shape descriptors than simple indicator functions---such as the skeletal density functions (SDF) shown in panel (c) \cite{Behandish2014a,Behandish2015,Behandish2015a,Behandish2016,Behandish2016a}. This enables assigning higher shape complementarity scores (thus lower haptic energies) to configurations with proper fit/contact, e.g., point E in panel (b). This comes at the expense of disappearance of the clear contact boundary between collision and no-collision regions, i.e., the $\conf-$obstacle, retrieved as the the $0-$sublevel set of the gap function $f^{-1}(0)$, and the free space.
\end{examp}

Additionally, the formulation of the energy function as a convolution allows using ideas from multivariate harmonic analysis \cite{Katznelson2004} to streamline haptic feedback computations. The convolution in the physical space (where the part geometries reside) transfers into a pointwise multiplication of the Fourier expansion of the SDFs for the individual parts (i.e., the `amplitudes' of the multi-dimensional SDF signals). Guided by this property, the formulation leads to a straightforward mathematical relationship between the Fourier representations of the SDF shape descriptors and the geometric energy field, which can benefit from the efficiency of the FFT algorithms \cite{Cooley1965}. Moreover, explicit analytic equations are obtained for computing the gradients of the convolution function (i.e., guidance forces and torques) for arbitrary spatial translations and rotations.

The performance is significantly improved using optimized FFT implementation on the highly-parallel GPU architecture. As a result, haptic-enabled simulation of realistic assembly scenarios with complex CAD models and low-clearance fits is made possible to an adequate fidelity with the application of GPU-accelerated FFT calls.

In addition to its theoretical generality, computational efficiency, and scalability with parallel-computing, the main advantages of this paradigm compared to the existing methods are the following:
\begin{enumerate}
    \item The analytic formulation is generic, allowing for a variety of different shape descriptors---ranging from simple PMC to intricate SDF---to be constructed using different kernels in the general formula.
    \item The decomposition of the shape descriptors into their Fourier components (i.e., the `modes' of the 3D signals) allows for a systematic means to trade off the accuracy of computations with the amount of available computation time and resources. In the case of haptic assembly, where there is a limited budget of time (ideally $1$ millisecond) available to each simulation frame, one could use truncated Fourier expansions (i.e., apply a `low-pass filter' to the 3D signals) to significantly speed up the convolution in real-time.
    \item The computational performance is indifferent to geometric and syntactic complexity of the objects in the physical domain. Unlike the existing combinatorial approaches to collision detection which scale in computation time with input complexity---e.g., the number of points, triangles, or voxels used in the representation---our method's efficiency depends solely on the degree of fidelity specified by the low-pass filter (i.e., number of retained dominant modes), and does not scale with the original input complexity.
\end{enumerate}

In spite of their promising theoretical and computational advantages, there are still important problems that need to be addressed and interesting research avenues to explore in order to realize the full potential of analytic methods:
\begin{itemize}
    \item The uniform sampling required by the classical uniform FFT algorithm \cite{Cooley1965} is not memory-efficient when large models with high levels are detail are rasterized. A promising new direction in this aspect is provided in \cite{Behandish2015d} using ideas from protein docking \cite{Bajaj2011} and tools such as the nonuniform FFT algorithm \cite{Potts2001}.
    \item Although the translational (i.e., commutative) component of the SDF convolution is handled efficiently by the Fourier methods, the rotational (i.e., noncommutative) component still requires sampling and interpolation. An interesting subject to look at in this area is the noncommutative harmonic analysis \cite{Chirikjian2010}.
    \item The preliminary validation results obtained so far are not nearly as strong as their counterparts presented in the literature for the currently popular approaches. In particular, the {\it scalability} of these methods from assembling pairs of rigid parts \cite{Behandish2014a,Behandish2015,Behandish2015a,Behandish2016,Behandish2016a} to gigantic assemblies with numerous rigid and flexible elements simulated at hierarchical levels of detail (LOD) is yet to be evaluated. Before this method can be adopted into physics simulation engines (PSE), robust implementations along with extensive testing are imperative.
\end{itemize}

\section{Conclusion} \label{sec_conclusion}

Haptic-enabled assembly planning has been restrained for a long time from achieving its full potential, due to the challenges presented by the competing objectives of handling high geometric complexity while maintaining a response rate of 1 kHz. The current computational models for constraint-based assembly guidance are either
\begin{enumerate}
    \item limited to the assembly of solids with very simple geometric features that are automatically detectable; or
    \item heavily dependent on user input for constraint specifications.
\end{enumerate}
Both methods generally presume {a priori} knowledge of the type of contact surfaces that one deals with, and are not generalizable to support objects of arbitrary shape. The majority of {ad hoc} solutions start from identifying the simplistic DOF-limiting constraints---e.g., restricting the contact features to planar, cylindrical, spherical, or conical surfaces or their intersection curves---followed by what can be conceptualized as simple energy formulations to enforce those constraints---e.g., spring-damper models to penalize the violation of co-planarity or co-axiallity conditions.

Lately, the dominant direction in this area has been aligned with a hybrid approach \cite{Vance2011,Perret2013}, separating the simulation into a `free motion' phase, using unilateral (i.e., inequality) `physical constraints' originated from collision detection; and a `fine insertion' phase, using bilateral (i.e., equality) `mating constraints' (e.g., of geometric and/or kinematic types) introduced artificially to limit the DOF. While the former fail to produce dynamically stable guidance for low-clearance insertion, the latter are either dependent on {a priori} manual specifications by the user, or are limited to simple semialgebraic features (e.g., planar, cylindrical, spherical, or conical) that can be identified automatically from CAD semantics using heuristic algorithms. The identification of the switch criteria between the two phases, on the one hand, and modeling the insertion constraints for different contact surface features, on the other hand, remained an open problem for objects of arbitrary shape.

An alternative promising direction is that of analytic methods \cite{Comba1968,Gilbert1985,Moreau1985} that have been popular in robotics for several decades and have found applications more recently in narrowphase CD \cite{Lysenko2013,Behandish2015d}. These methods are advantageous over combinatorial techniques in their indifference to topological, geometric, and syntactic complexities, as well as the meaningful trade-off mechanism they provide between accuracy and performance for time-critical applications ``on a budget'' by Fourier filtering. An extension of the Fourier CD method is only recently being applied to haptic assembly \cite{Behandish2014a,Behandish2015,Behandish2015a,Behandish2016,Behandish2016a}, which formulates a `geometric energy' field to unify the collision response and geometric guidance into a single interaction and applies to arbitrarily complex shapes. Although this unified paradigm provides a promising alternative direction for haptic assembly, its practical application to real-world assembly planning, simulation, and verification scenarios at an industrial capacity is yet to be tested, which, in turn, is contingent upon its integration into existing PSEs.

\section{Acknowledgements}

This work was supported in part by the National Science Foundation grants CMMI-1200089, CMMI-1462759, and CISE-1526249.

\appendix

\section{Literature Catalogue} \label{app_catalogue}

In this appendix we present examples of several important and relatively recent studies and systems that use haptic support for a variety of VP/VA activities, catalogued on the basis of authoring research groups.

Different applications of haptics to design, analysis, validation, and manufacture process planning are reviewed in Section \ref{app_VP}. The applications of haptics to assembly and disassembly related problems are exclusively reviewed in greater extent in Section \ref{app_VA}.
The enumerated collection is by no measure comprehensive, but presents a broad selection of a range of different research developments---the most complete collection in one place to the present date, to the best of our knowledge.

\subsection{Haptic-Enabled VP} \label{app_VP}

Hollerbach et al. at the University of Utah developed a `haptic display' for grasping and manipulating virtual mechanisms (e.g., linkages and chains) \cite{Maekawa1998,Nahvi1998} using an exoskeleton haptic device \textsf{Sarcos Dextrous Arm Master}---later upgraded to \textsf{Sarcos DTS Master Exoskeleton} for subsequent work \cite{Frey2008}. They also integrated the haptic interface with Utah's \textsf{Alpha-1} geometric modeling system to enable manipulation of both polygonal (i.e., mesh) and freeform (i.e., parametric) surfaces, particularly using direct parametric tracing (DPT) \cite{Thompson1997a} for tracing untrimmed and trimmed NURBS surfaces \cite{Thompson1997,Thompson2005} and physics-based models---e.g., stick-slip friction \cite{Salcudean1997} and nonlinear viscosity \cite{Marhefka1996}---for rapid virtual prototyping \cite{Hollerbach1997,Hollerbach1996}. Among other related works of the group is nonlinear device modeling for VR applications \cite{Colton2004,Colton2005}.

\begin{table*}
\centering
\caption{A chronological review of haptic-enabled virtual prototyping platforms for various product development tasks.}  \label{tab_VPlit}
\begin{adjustbox}{max width=0.95\textwidth}
\begin{threeparttable}
    \vspace{-0.2cm}
    \begin{tabular}{l  l  l  l  l  l  l}
        \toprule
        \textbf{References} & \textbf{Year} & \textbf{System} & \textbf{Methods} & \textbf{Software} & \textbf{Hardware} & \textbf{Key Features} \\
        \midrule
        \cite{Hollerbach1996} & 1996 & --- & PBM, & Utah's \textsf{Alpha-1}, & \textsf{Sarcos Dextrous} & exoskeleton haptic interface for CAD;\\
        \cite{Hollerbach1997,Thompson1997,Thompson1997a}   & 1997 & & CD.  & \textsf{ControlShell},   & \textsf{Arm Master}. & uses mesh and DPT for elastic CD;\\
        \cite{Maekawa1998,Nahvi1998}         & 1998 & &      & TCP and UDP.             &                      & asynchronous networking with device.\\
        \cite{Thompson2005}                  & 2005 & &      & &                      & \\
        \midrule
        \cite{Dachille1999}   & 1999 & --- & PBM, & \textsf{GHOST}$^\circledR$ API. & \textsf{Phantom}$^\circledR$ \textsf{1.0}. & freeform sculpting using B-splines;\\
        \cite{Dachille2001}   & 2001 & & MSS, &  &  & mass-spring discretization of surfaces;\\
                              &      & & CD.  &  &  & dynamic optimization of control points.\\
        \midrule
        \cite{McDonnell2001a,McDonnell2001} & 2001 & \textsf{Digital} & PBM, & \textsf{GHOST}$^\circledR$ API. & \textsf{Phantom}$^\circledR$ \textsf{1.0}. & `virtual clay' using subdivision solids;\\
        \cite{Chang2002,McDonnell2002}      & 2002 & \textsf{Sculpture} & DSS, &  &  & mass-spring discretization of lattice;\\
        \cite{McDonnell2005}                & 2005 & & CD.  &  &  & dynamic optimization of mass points.\\
        \cite{McDonnell2007}                & 2007 & &      &  &  & \\
        \midrule
        \cite{Evans2000} & 2000 & --- & PBM, & \textsf{FreeForm}$^\circledR$. & \textsf{Phantom}$^\circledR$ \textsf{Desktop}$^\circledR$. & concept generation via sketch elevation;\\
        \cite{Cheshire2001} & 2001 & & CD.  & & & form approximation; surface shaping;\\
        \cite{Evans2005}    & 2005 & &      & & & wire-cutting, smoothing, and mirroring.\\
        \midrule
        \cite{Liu2003}      & 2003 & \textsf{Virtual} & CBM, &  \textsf{GHOST}$^\circledR$ API, & \textsf{Phantom}$^\circledR$ \textsf{Desktop}$^\circledR$, & \textsf{COM}-based CAD-VR interoperability; \\
        \cite{Liu2004,Liu2004a} & 2004 & \textsf{DesignWorks} & CD.  &  \textsf{MS} \textsf{COM}+. & Unspecified OST-HMD, & freeform-based B-rep surface operations;\\
        \cite{Liu2005}      & 2005 & &      & & \textsf{5DT FOB}$^\circledR$ OTs. & shape control functions and SVD.\\
        \midrule
        \cite{Yang2003}  & 2003 & --- & PBM, &  \textsf{FreeForm}$^\circledR$, & \textsf{Phantom}$^\circledR$ \textsf{Desktop}$^\circledR$, & uses mass-spring model for elastic CD;\\
        \cite{Chen2004,Chen2004b,Yang2004}     & 2004 &                & RLE, &  \textsf{GHOST}$^\circledR$ API,  &  \textsf{ABB IRB 1400}. & uses S-RLE description for plastic CD;\\
        \cite{Chen2005a,Yang2005,Yang2005a}    & 2005 &                & FEM, &  \textsf{VTK} Toolkit. &  & both additive and subtractive forming;\\
        \cite{Chen2007}                        & 2007 &                & CD.  &                        &  & models milling and path generation.\\
        \midrule
        \cite{Chen2004a}              & 2004 & \textsf{HVCMM} & PBM, &  \textsf{GHOST}$^\circledR$ API,  &  \textsf{Phantom}$^\circledR$ \textsf{Desktop}$^\circledR$. & uses S-RLE and custom model for CD;\\
        \cite{Chen2005b,Yang2005b}    & 2005 &                & RLE, &  &  & models CMM inspection path planning;\\
        \cite{Wang2006a}              & 2006 &                & CD.  &  &  & models CMM accessibility analysis.\\
        \cite{Wang2009a}              & 2009 &                &      &  &  & \\
        \midrule
        \cite{Gao2004}        & 2004 & --- & PBM, & \textsf{GHOST}$^\circledR$ API. & \textsf{Phantom}$^\circledR$ \textsf{Premium}$^\circledR$. & freeform sculpting using B-splines;\\
        \cite{Gao2005,Gao2005a} & 2005 & & MSS, &  &  & mass-spring discretization of surfaces;\\
        \cite{Gao2006,Gao2006a} & 2006 & & CD.  &  &  & dynamic optimization of control points;\\
        \cite{Gao2007}        & 2007 & &      &  &  & implicit modeling of prob/tool heads;\\
        \midrule
        \cite{Convard2004}    & 2004 & \textsf{VRAD}     & PBM, & \textsf{EVI3d} Drivers, & \textsf{CAVE}-like System, & allows implicit edition of CHG data;\\
        \cite{Picon2008,Picon2008a,Picon2008b} & 2008 &  & CD.  & \textsf{VEserver},    & head tracking devices, & models haptic selection and extrusion;\\
        \cite{Simard2009}     & 2009 &                   &      & \textsf{OpenCASCADE}, & WTP haptic devices, & enables multimodal interactions;\\
        \cite{Bourdot2010}    & 2010 &                   &      & \textsf{OpenGL}. & \textsf{IBM ViaVoice}. & future developments aimed at \textsf{CATIA}$^\circledR$.\\
        \midrule
        \cite{Bordegoni2004}  & 2004 & \textsf{T'nD}  & PBM, & Device APIs. & \textsf{FCS-HapticMaster}, & uses tesselated models for CD;\\
        \cite{Bordegoni2006,Bordegoni2006b,Bordegoni2006a}  & 2006 & & CD.  & & \textsf{nVisor ST} OST-HMD, & uses chip removal theory for PBM;\\
        \cite{Cugini2007}     & 2007 &                &      & & \textsf{Vicon M2-460}.    & conceptual design and ergonomics;\\
        \cite{Bordegoni2010}  & 2010 &                &      & &                           & creates NURBS for downstream.\\
        \midrule
        \cite{Bordegoni2004}  & 2004 & \textsf{VeRVE}    & PBM. & Device APIs, & Haptic Knob(s), & uses `smart' haptic feedback (tacton);\\
        \cite{Bordegoni2006}  & 2006 &                   &      & \textsf{UGS Jack}. & \textsf{nVisor ST} OST-HMD, & models ergonomic interactive testing.\\
        \cite{Bordegoni2007a,Bordegoni2007} & 2007   &   &      & & \textsf{Vicon M2-460}. &\\
        \midrule
        \cite{Bordegoni2008a} & 2008 & \textsf{PROGIMM}  & PBM, & \textsf{3DVIA Virtools}, & \textsf{KUKA} \textsf{KRC} and \textsf{KR3}, & mixed reality and mixed prototyping;\\
        \cite{Bordegoni2009a,Bordegoni2009b}  & 2009 &   & CD.  & \textsf{KRL}$^\textrm{TM}$+XML. & \textsf{nVisor ST} OST-HMD,    & ergonomic assessment of driver seats;\\
        \cite{Caruso2010}                     & 2010 &   &      & & \textsf{Vicon M2-460}. & haptic tools for automotive industry. \\
        \cite{Caruso2011a,Caruso2011}         & 2011 &   &      & &                        & \\
        \midrule
        \cite{Bordegoni2008}  & 2008 & \textsf{SATIN}   & PBM, & \textsf{SML} Manager, & \textsf{FCS-HapticMaster}$\times 2$, & haptic strip for curve approximation;\\
        \cite{Bordegoni2010a,Ferrise2010,Araujo2010} & 2010 &  & CD.  & \textsf{ThinkCore} API, & \textsf{DVHDS} Components, & various haptic modules and knobs;\\
        \cite{Bordegoni2012}  & 2012 &                   &      & \textsf{OpenIVI}, & \textsf{nVisor ST} OST-HMD, & CAD modeling and shape analysis;\\
                              &      &                   &      & TCP/IP. & \textsf{Vicon M2-460}. & conceptual and aesthetic design.\\
        \midrule
        \cite{Liu2012,Liu2012}      & 2012 & \textsf{DesignWorks} & CBM, & \textsf{GHOST}$^\circledR$ API, & \textsf{Phantom}$^\circledR$ \textsf{Desktop}$^\circledR$, & variational B-spline editing techniques;\\
        \cite{Liu2014}      & 2014 & & CD.  & \textsf{MS} \textsf{COM}+, & Unspecified HMD, & real-time energy minimization.\\
        \cite{Liu2015}      & 2015 & &      & \textsf{VBS} Kernel. & \textsf{5DT FOB}$^\circledR$ OTs. & \\
        \bottomrule
    \end{tabular}
    \begin{tablenotes}
        \footnotesize
        \item \textbf{Abbreviations used for generic terms:} physically-based modeling (PBM), constraint-based modeling (CBM), collision detection (CD), virtual environment (VE), virtual reality (VR, augmented reality (AR), computer-aided design (CAD), application programmer's interface (API), software development kit (SDK), transmission control protocol (TCP), user datagram protocol (UDP), internet protocol (IP), direct parametric tracing (DPT), mass-spring system (MSS), dynamic subdivision solids (DSS), optical see-through (OST), head-mounted display (HMD), boundary representation (B-rep), optical tracker (OT), singular value decomposition (SVD), run-length encoding (RLE), finite element method (FEM), spatial run-length encoding (S-RLE), coordinate measuring machine (CMM), wand-type pointer (WTP), construction history graph (CHG), nonuniform rational B-splines (NURBS), extensible markup language (XML).
        \item \textbf{Abbreviations used for brand names:} \textsf{general haptic open software toolkit (GHOST), haptic virtual coordinate measuring machine (HVCMM), virtual reality aided design (VRAD), cave automatic virtual environment (CAVE), touch and design (T'nD), virtual reality system for validation of equipment controls (VeRVE), la progettazione immersiva multisensoriale (PROGIMM), sound and tangible interfaces for novel product design (SATIN), Microsoft (MS), Component Object Model (COM), Flock of Birds (FOB).}
        \item {}{$\times2$} means two-handed haptic device or a pair of one-handed devices.
    \end{tablenotes}
\end{threeparttable}
\end{adjustbox}
\end{table*}

Qin et al. at the State University of New York at Stony Brook developed a variety of haptic sculpting systems by applying physics-based modeling based on lumped mass-spring networks---made of `control points' and `mass points' connected within a control mesh and a network of springs---and Newtonian dynamics to different solid representations. The representations include B-spline surfaces discretized with linear springs over the control mesh \cite{Dachille2001,Dachille1999}; dynamic subdivision solids \cite{McDonnell2000} discretized with both linear and angular springs over the control lattice (called the `virtual clay' method)%
\footnote{Rossignac et al. at the Georgia Institute of Technology used a similar project name (the `digital clay' project) \cite{Gargus2002,Rossignac2003,Bosscher2003} for the development of a new type of haptic interface for finger sculpting; namely, a computer-controlled physical surface that deforms in response to the pressure changes exerted by bare hands \cite{Gargus2002,Rossignac2003}, built using `formable crust architectures' \cite{Bosscher2003}. Ishii et al. at the MIT Media Lab recently developed a similar concept called \textsf{InForm} \cite{Leithinger2014,Leithinger2015} for `physical telepresence' and remote collaboration.}
\cite{McDonnell2002,McDonnell2001a,McDonnell2001}; volumetric implicit functions (i.e., `density fields') used to define semialgebraic sets bounded by a finite number of B-spline patches discretized into a grid of `density springs' \cite{Hua2002,Hua2002a,Hua2004}---which are also capable of performing CSG operations and knot insertion; and dynamic pointset surfaces by fitting implicit functions to local distance fields and applying ideas from implicit modeling \cite{Guo2004}. The group later implemented the ideas from both volumetric subdivisions and implicit modeling into a system called \textsf{DigitalSculpture} \cite{McDonnell2005} for interactive surface editing. Among their other relevant works are direct mesh editing using PDE-based geometric surface flow in a system called \textsf{HapticFlow} \cite{Duan2004} and applying FEM to incorporate flexibility into subdivision solid geometry for haptic sculpting \cite{McDonnell2007}. A \textsf{Phantom}$^\circledR$ \textsf{1.0} device (3 DOF input, 3 DOF output) was used for all applications.

Liu et al. form the Queen's University of Belfast developed the first VR-CAD system called \textsf{Virtual DesignWorks} \cite{Liu2003,Liu2004} that used \textsf{Microsoft}' \textsf{COM+} technology  for real-time exchange of models between a CAD module (e.g., NURBS-based B-reps for flexible editing) and a haptic module (e.g., polygon/voxel-based for fast rendering) for freeform surface editions (e.g., pulling, pushing, and dragging). The \textsf{COM}-based implementation enabled real-time interoperability between a typical CAD kernel (e.g., used in \textsf{SolidWorks}$^\circledR$, \textsf{Unigraphics}$^\circledR$, etc.) and an approximate representation of geometry for haptics. In a follow-up study \cite{Liu2004a,Liu2004} they used `shape control functions' to simulate surface deformations which yield a linear system of constraints solved by the singular value decomposition (SVD) method. Later in Hebei University of Technology, Liu also implemented variational B-spline (VBS) techniques and real-time energy minimization (using Hebei's \textsf{VBS} kernel) for surface hole filling \cite{Liu2012a,Liu2015} and for interactive surface editing \cite{Liu2012,Liu2014} into \textsf{DesignWorks}.%
\footnote{The later versions of \textsf{Virtual DesignWorks} \cite{Liu2003,Liu2004} was called \textsf{DesignWorks} \cite{Liu2012,Liu2014} in subsequent publications.}
A \textsf{Phantom}$^\circledR$ \textsf{Desktop}$^\circledR$ device (6 DOF input, 3 DOF output) was used for all applications.

Chen et al. at the University of Hong Kong developed a product development platform with a wide range of haptic functionalities \cite{Chen2007,Chen2004b,Chen2005a} including machine tool path planning \cite{Yang2003}---e.g., for 5-axis milling based on the method used in \textsf{MIT Suzuki} haptic system \cite{Balasubramaniam2002}; real-time mechanical property analysis \cite{Yang2005}---e.g., using a hierarchical finite element method (FEM) from \cite{Kang1996}; reverse engineering and shape digitizing \cite{Yang2004,Yang2005a}---e.g., using haptic-guided volume sculpting method from \cite{Iwata1993}; and a module called \textsf{HVCMM} \cite{Chen2004a,Chen2005b,Yang2005b,Wang2006a,Wang2009a} for inspection path generation for coordinate measurement machines (CMM). The platform uses their own volumetric enumeration data structure called spatial run-length encoding (S-RLE) \cite{Chen2004} for geometric rasterization and haptic rendering. The group's subsequent works also include haptic-guided repair of triangular meshes (e.g., hole-filling) \cite{He2006}, haptic-assisted evaluation of compliant mechanisms \cite{Tang2007}, and surface texture and friction modeling for tactile feedback \cite{Chen2008}. A \textsf{Phantom}$^\circledR$ \textsf{Desktop}$^\circledR$ device (6 DOF input, 3 DOF output) was used for all applications.

Gao et al. form the University of Hong Kong and Gibson et al. from the National University of Singapore developed a haptic sketching system for manipulating 3D B-spline curves \cite{Gao2004} and a haptic sculpting system \cite{Gao2005,Gao2005a,Gao2006} to create and modify B-spline surfaces using a variety of prob/tool head geometries. The sculpting system enables intuitive pushing and pulling operations on freeform surfaces to relate the virtual modeling experience to the physical world experience. The model uses a mass-spring discretization of the surface checked against the implicit representation of the prob profile for elastic collision response. The implementation also enables wavelet-based multiresolution representations, which in turn enables sweep editing and 3D texture reuse in the frequency domain \cite{Gao2006}. The system was later added with functionalities to design and evaluate multimaterial products \cite{Gao2006a} and to work with point clouds and NURBS \cite{Gao2007}. A \textsf{Phantom}$^\circledR$ \textsf{Premium}$^\circledR$ device (6 DOF input, 6 DOF output) was used for all applications.

The VENISE research team (Bourdot et al.) at the CNRS/LIMSI laboratory in partnership with Universite Paris-Sud developed an integrated and immersive VR-CAD system called \textsf{VRAD} \cite{Bourdot2010} that supports modification of CAD semantics directly within the VE. The system enables direct modification of construction history graphs (CHG) by selection of primitive elements on the evaluated B-reps \cite{Convard2004}. The haptic functionality \cite{Picon2008} was added using a variety of force models---namely, spring elasticity and attraction force models for haptic selection \cite{Picon2008a}, `reference-based', `haptic-grid', and graduated virtual fixtures (GVF) methods for haptic extrusion \cite{Picon2008b}, and potential field approaches for haptic selection \cite{Simard2009}. The implementation supports multimodal interactions using previous tools developed by the group such as \textsf{EVI3d} \cite{Bourdot1998,Touraine2002} including multimodal `fusion' using their own \textsf{VEserver} \cite{Touraine2001} concurrently running on multiple computers, and multimodal `fission' for managing large and complex data in future work \cite{Bouyer2007}. Among other works of the group is flexible model rendering \cite{Convard2005}. The device brands were not specified.

The KAEMaRT research group (Cugini et al. and Bordegoni et al.) at the Politecnico di Milano developed a system called \textsf{T'nD} \cite{Bordegoni2006,Bordegoni2006b,Bordegoni2010} to perform conceptual design activities using virtual tools attached to a 6DOF \textsf{FCS-HapticMaster} device to simulate physical form-making activities---e.g., surface scraping using rakes \cite{Bordegoni2006a} and finishing using sandpaper \cite{Cugini2007} modeled using chip removal theory \cite{Merchant1945}. The group also developed other prototypes such as \textsf{VeRVE} \cite{Bordegoni2006,Bordegoni2007a,Bordegoni2007} for ergonomic validation using haptic knobs \cite{Colombo2006}, and \textsf{PROGIMM} \cite{Bordegoni2008a,Bordegoni2009a,Bordegoni2009b} as a mixed-reality platform for validating prototypes in the automotive industry in collaboration with Caruso et al. \cite{Caruso2010,Caruso2011a,Caruso2011}. These conceptual design and ergonomic validation functionalities were also integrated, in addition to aesthetic design tools \cite{Bordegoni2008,Bordegoni2010a,Bordegoni2012} into a multimodal and multisensory system called \textsf{SATIN} \cite{Ferrise2010,Araujo2010} using a deformable strip attached to two \textsf{FCS-HapticMaster} devices installed in a parallel configuration and a display system called \textsf{DVHDS} composed of projectors and mirrors for superimposing the virtual scene over the user's hands.

Table \ref{tab_VPlit} presents a more inclusive list (including studies and systems not described above) and a summary of their key features.

\subsection{Haptic-Enabled VA} \label{app_VA}

Gupta et al. at the MIT Media Lab developed one of the earliest multimodal desktop VA systems (i.e., with visual, auditory, and touch feedback) called \textsf{VEDA} \cite{Gupta1995,Gupta1997a,Gupta1997} that used physically-based modeling (PBM) for part behavior, with the end-goal of integrating design evaluation techniques such as design for assembly (DFA) into the existing CAD systems. The shapes were limited to either convex or concave rigid 2D polygons whose contact was modeled using Coulomb's laws of (static and dynamic) friction. Dual \textsf{Phantom}$^\circledR$ \textsf{1.0} devices (6 DOF input, 3 DOF output) were used. The system's performance was evaluated for simple peg-in-hole examples.

Jayaram et al. at the Washington State University in a research partnership with Lyons and Hart from the National Institute of Standards and Technology (NIST) developed the immersive assembly planning and evaluation platform called \textsf{VADE} \cite{Jayaram1997,Jayaram1999,Jayaram1999a}. The system directly imported CAD models from \textsf{Pro/ENGINEER}$^\circledR$ and performed assembly using constraint-based modeling (CBM) \cite{Wang2003} in addition to some basic PBM. It was further augmented with additional functionalities including interactive swept volume generation and modification using synchronous links between the VE (providing trajectory data) and the CAD (providing shape data) sub-systems \cite{Yang1999}, inclusion of hand-held tools (e.g., screwdrivers) and the corresponding assembly operations and tool/hand/part interactions \cite{Jayaram2000}, and distributed environments for collaborative assembly \cite{Jayaram2000a}. Additionally, workplace ergonomic evaluation tools---e.g., the rapid upper-limb assessment (RULA) algorithm \cite{McAtamney1993}---were integrated later into the system \cite{Shaikh2004,Shaikh2003}. Subsequent industrial case studies \cite{Jayaram2004,Jayaram2007} concluded the feasibility of VA methods for deployment in the actual PLM cycle, and identified the key issues to be resolved in terms of ease of use, portability of the applications, and preparation of the evaluation models. Both one- and two-handed assembly was enabled using \textsf{CyberGrasp}$^\circledR$ haptic gloves (5 DOF input, 5 DOF output).

Bras et al. at the Georgia Tech. developed another early haptic-enabled assembly and disassembly simulation environments called \textsf{HIDRA} \cite{McDermott1999,Coutee2001,Coutee2002}. The implementation used two concurrent loops (one for graphics and one for haptics+CD) to comply with the difference in frame rate requirements. To allow for fast CD, they used University of Minnesota's \textsf{Qhull} library for computing the convex hull of parts (or an obvious decomposition of the parts). For the CD itself between convex elements, Mitsubishi Electric Research Laboratory (MERL)'s \textsf{V-Clip} library \cite{Mirtich1998} was used in the earlier versions of \textsf{HIDRA} \cite{McDermott1999,Coutee2001}. Later, to enable faster multibody processing, sweep generation, and prune sorting, the use of \textsf{SWIFT} \cite{Ehmann2000} and \textsf{SWIFT++} \cite{Ehmann2001} was investigated in \cite{Coutee2002}, where the \textsf{++} version enables nonconvex CD by decomposing into convex elements. The study discusses optimization techniques such as part anchoring, dynamic loading of haptic representations, and partitioning of part updates to minimize the effects of inherent computational limitations. The results of the study indicated limited usability of haptics for simple peg-in-hole examples. A \textsf{Phantom}$^\circledR$ \textsf{1.0} device (6 DOF input, 3 DOF output) was used.

Borro et al. at the CEIT research group (CEIT-IK4 research alliance as of 2004) developed the system called \textsf{REVIMA} \cite{Savall2002,Borro2004} for haptic-assisted maintainability simulation in aeronautics (e.g., assembling aircraft engine mock-ups). A specialized large haptic device called \textsf{LHIfAM} (6 DOF input, 3 DOF output) was developed for this purpose.

Wan et al. at the Zhejiang University developed a multimodal VE system for assembly named \textsf{MIVAS} \cite{Wan2004,Zhu2004} with voice input, stereo visual feedback (\textsf{CAVE}-like system), sound feedback, and force feedback. To facilitate an automatic translation of the assembly model from \textsf{Pro/ENGINEER}$^\circledR$ CAD models, an interface was developed using \textsf{Pro/TOOLKIT}$^\circledR$ which automatically imports geometry and topology (for PBM) as well as assembly constraints (for CBM) into the VE. It also incorporated models of virtual hand kinematics and grasping heuristic patterns for realistic user interaction. Virtual hand/part CD for grasp feedback was carried out using \textsf{RAPID} \cite{Gottschalk1996} while fast part/part CD was implemented using \textsf{VPS}$^\text{TM}$ \cite{Wan2003,McNeely2005a,McNeely2005,McNeely2006}. \textsf{CyberGrasp}$^\circledR$ haptic gloves (5 DOF input, 5 DOF output) were used to enable force feedback from the virtual hand.

\begin{table*}
\centering
\caption{A chronological review of haptic-enabled virtual prototyping platforms for assembly and disassembly tasks.} \label{tab_VAlit}
\begin{adjustbox}{max width=0.95\textwidth}
\begin{threeparttable}
    \vspace{-0.2cm}
    \begin{tabular}{l  l  l  l  l  l  l}
        \toprule
        \textbf{References} & \textbf{Year} & \textbf{System} & \textbf{Methods} & \textbf{Software} & \textbf{Hardware} & \textbf{Key Features} \\
        \midrule
        \cite{Gupta1995}               & 1995 & \textsf{VEDA}  & PBM, &  \textsf{GHOST}$^\circledR$ API, & \textsf{Phantom}$^\circledR$ \textsf{1.0}$\times 2$, & `vertex-edge' pair contact modeling;\\
        \cite{Gupta1997a,Gupta1997}    & 1997 &                & CD.  &  & \textsf{SGI}$^\circledR$ \textsf{Indigo}$^\text{TM}$. & uses Coulomb's law of friction;\\
                            &   &                 &      &  &  & limited to 2D convex/concave polygons.\\
        \midrule
        \cite{Jayaram1997}  & 1997 & \textsf{VADE}   & PBM,      & Device APIs. & \textsf{CyberGrasp}$^\circledR$$\times 2$, & one- \& two-handed dexterous assembly;\\
        \cite{Jayaram1999,Jayaram1999a,Yang1999} & 1999 & & CBM, &   & \textsf{SGI}$^\circledR$ \textsf{Onyx2}$^\text{TM}$, & approximate line-polygon intersections;\\
        \cite{Jayaram2000,Jayaram2000a}  & 2000 &    & CD.       &   & \textsf{InfiniteReality} Pipes, & maintains link between VE and CAD;\\
        \cite{Jayaram2001,Wang2001}      & 2001 &    &           &   & \textsf{5DT FOB}$^\circledR$ OTs, & imports ``assembly intent'' from CAD;\\
        \cite{Shaikh2003,Wang2003}       & 2003 &    &           &   & Unspecified HMD, & sweep generation and trajectory editing;\\
        \cite{Jayaram2001,Shaikh2004}    & 2004 &    &           &   & \textsf{ImmersaDesk}$^\text{TM}$. & simulates fasteners and screwdrivers;\\
        \cite{Jayaram2006}               & 2006 &    &           &   &  & ergonomic evaluation tools;\\
        \cite{Jayaram2007}               & 2007 &    &           &   &  & collaborative/distributed assembly.\\
        \midrule
        \cite{McDermott1999}   & 1999 & \textsf{HIDRA}  & CD. &  \textsf{GHOST}$^\circledR$ API, & \textsf{Phantom}$^\circledR$ \textsf{1.0}. & CD-based assembly and disassembly;\\
        \cite{Coutee2001}   & 2001 &                 &      &  \textsf{OpenGL}, \textsf{Motif}$^\text{TM}$ &  & uses \textsf{Qhull} for convex hull generation;\\
        \cite{Coutee2002}   & 2002 &                 &      &  \textsf{Qhull}, \textsf{V-Clip}, &  & uses \textsf{V-Clip} and \textsf{SWIFT(++)} for CD;\\
                            &      &                 &      &  \textsf{SWIFT(++)}. &  & limited use to simple peg-in-holes.\\
        \midrule
        \cite{Savall2002}   & 2002 & \textsf{REVIMA} & PBM, &  Device APIs, & CEIT \textsf{LHIfAM}. & maintenance of aircraft engines;\\
        \cite{Borro2004}    & 2004 &                 & CD.  &  \textsf{OpenGL}. & & accessibility and interference analysis.\\
        \midrule
        \cite{Wan2004,Zhu2004} & 2004 & \textsf{MIVAS}  & PBM, & \textsf{OpenGL Optimizer},  & \textsf{CyberGrasp}$^\circledR$, & multimodal VE for grasp+assembly; \\
                               &      &                 & CBM, & \textsf{Pro/TOOLKIT}$^\circledR$  & \textsf{SGI}$^\circledR$ \textsf{Onyx2}$^\text{TM}$, & direct constraint import from CAD; \\
                               &      &                 & CD.  & \textsf{IBM ViaVoice}, & \textsf{CAVE}-like System & allows assembly sequence generation; \\
                               &      &                 &      & TCP/IP. & \textsf{CrystalEye} SG, & allows assembly trajectory generation; \\
                               &      &                 &      &         & \textsf{5DT FOB}$^\circledR$. & optimization techniques for assembly. \\
        \midrule
        \cite{Seth2005} & 2005 & \textsf{SHARP}  & PBM, & \textsf{VR Juggler},  & \textsf{Phantom}$^\circledR$ \textsf{Omni}$^\circledR$$\times 2$, & uses \textsf{VPS}$^{\text{TM}}$ for high-clearance CD; \\
        \cite{Seth2006} & 2006 &                 & CBM, & \textsf{OpenGL PSG},  & \textsf{Barco Baron} PT, & uses \textsf{D-Cubed} for low-clearance CD;\\
        \cite{Seth2007} & 2007 &                 & CD.  & \textsf{OpenHaptics}$^\circledR$, & \textsf{CAVE}-like System, & swept volumes for maintainability; \\
        \cite{Seth2008} & 2008 &                 &          & \textsf{VPS}$^{\text{TM}}$, \textsf{D-Cubed}, & Unspecified HMD.   &  capable of creating subassemblies; \\
        \cite{Seth2010} & 2010 &                 &          & TCP/IP.      & &  capable of networking with others. \\
        \midrule
        \cite{Lim2006,Lim2006a} & 2006 & \textsf{HAM(M)S}  & PBM, & \textsf{OpenHaptics}$^\circledR$,  & \textsf{Phantom}$^\circledR$ \textsf{Omni}$^\circledR$$\times 2$, & TCT evaluations for performance; \\
        \cite{Lim2007a,Lim2007} & 2007 &                 & CD.  & \textsf{VTK} Toolkit,  & \textsf{Phantom}$^\circledR$ \textsf{Desktop}$^\circledR$. & EMG evaluations for motor control; \\
        \cite{Lim2010}          & 2010 &                 &      & \textsf{PhysX}$^{\text{TM}}$ SDK, & & MTL \& therblig analysis of motion; \\
        \cite{Badillo2013}      & 2013 &                 &      & \textsf{Bullet Physics}, & & multiple PSEs for PBM+CD; \\
        \cite{Badillo2014,Badillo2014a,Badillo2014b}      & 2014 &  &      & \textsf{MFC} library. & & assembly process modeling \& planning. \\
        \midrule
        \cite{Iglesias2006,Iglesias2006a}    & 2006 & \textsf{(C)HAS} & CBM, &  \textsf{OpenHaptics}$^\circledR$, & \textsf{Phantom}$^\circledR$ \textsf{Omni}$^\circledR$, & distributed \& collaborative assembly;\\
        \cite{Iglesias2007}    & 2006 & & CD.  &  Labein's \textsf{DATum}, & \textsf{Phantom}$^\circledR$ \textsf{Premium}$^\circledR$, & combines assembly \& haptic simulators;\\
        \cite{Iglesias2008}    & 2006 & &      &  \textsf{RAPID}.        & \textsf{PERCRO} \textsf{GRAB}. & tested on an aeronautical assembly.\\
        \midrule
        \cite{Iacob2007}    & 2007 & \textsf{CVE}   & PBM, & \textsf{VirtuoseAPI}, & \textsf{Virtuose6D}$^\text{TM}$ \textsf{35-45}, & collaborative VE; modular behavior;\\
        \cite{Iacob2008}    & 2008 &                & CBM, & \textsf{VTK} Toolkit, & 2D Wall Display, & automatic contact constraint detection;\\
        \cite{Iacob2011}    & 2011 &                & CD.  & \textsf{ODE}. & Crystal Eye SG$^\ast$, & mobility trajectory characterization;\\
        \cite{Iacob2012}    & 2012 &                &      &  & Christie HD3 SP. & classifies simple `functional surfaces';\\
        \cite{Iacob2013}    & 2013 &                &      &  &  & uses dual-quaternion representation.\\
        \midrule
        \cite{Bhatti2008}   & 2008 & \textsf{HIIVR} & PBM, & Device APIs, & \textsf{Phantom}$^\circledR$ \textsf{Omni}$^\circledR$, & used for procedural skills development;\\
        \cite{Bhatti2009,Jia2009}  & 2009 &         & CD.  & \textsf{SmartCollisions}$^\circledR$. & \textsf{5DT FOB}$^\circledR$ \& DGs, & different difficulty levels for training;\\
                            &      &                &      &  & \textsf{eMagin}$^\circledR$ \textsf{Z800} HMD, & assists the user by visual cues;\\
                            &      &                &      &  & \textsf{NEC}$^\circledR$ SP. & evaluated using SE \& PVE scales;\\
        \midrule
        \cite{Christiand2008} & 2008 & --- & CBM, &  \textsf{CHAI3D}, & \textsf{Phantom}$^\circledR$ \textsf{Omni}$^\circledR$. & optimization of path \& sequence;\\
        \cite{Christiand2009} & 2009 &     & CD.  & & & AABB CD for 2D polygonal shapes;\\
        \cite{Christiand2011} & 2011 &     &      & & & improved assembly (time \& distance).\\
        \midrule
        \cite{Garbaya2009,Zaldivar-Colado2009}   & 2009 & \textsf{VEDAP-II} & PBM, &  \textsf{VHT} Toolkit, & \textsf{CyberGlove}$^\circledR$, & models grasp+move+locate+secure;\\
                               &      & & CD. &  \textsf{PhysX}$^\text{TM}$ SDK, & \textsf{CyberGrasp}$^\circledR$, & focuses on virtual coupling dynamics;\\
                               &      & &     &  \textsf{OpenGL}, \textsf{V-Clip}. & \textsf{CyberForce}$^\circledR$, & models `visual dynamic behavior'.\\
        \midrule
        \cite{Bordegoni2009}   & 2009 & --- & PBM, &  \textsf{VirtuoseAPI}, & \textsf{Virtuose6D}$^\text{TM}$ \textsf{35-45}, & evaluation of two-handed assembly;\\
                        &      &        & CD.  &  \textsf{Cyviz Viz3D}$^\text{TM}$. & \textsf{WiiRemote}$^\text{TM}$, & heuristic criteria for quality assessment;\\
                        &      &        &      &  \textsf{Dassault}$^\circledR$ \textsf{Virtools}, & Unspecified OTs. & low- vs. high-cost device assessment.\\
                        &      &        &      &  \textsf{ARTrack}$^\text{TM}$, \textsf{IPP}. &  & real-scale projection and tracking.\\
        \midrule
        \cite{Stoll2009}    & 2009 & --- & CD.  & \textsf{Novint SDK}, & \textsf{Novint} \textsf{Falcon}$^\circledR$ & telepresence for in-space assembly;\\
                            &      &     &      &  &  & uses `virtual walls/boundaries' for CD;\\
                            &      &     &      &  &  & tested on \textsf{NASA}'s \textsf{SPHERES} testbed.\\
        \bottomrule
    \end{tabular}
\end{threeparttable}
\end{adjustbox}
\end{table*}
\begin{table*}
\centering
\begin{adjustbox}{max width=0.95\textwidth}
\begin{threeparttable}
    \vspace{-0.2cm}
    \begin{tabular}{l  l  l  l  l  l  l}
        \toprule
        \textbf{References} & \textbf{Year} & \textbf{System} & \textbf{Methods} & \textbf{Software} & \textbf{Hardware} & \textbf{Key Features} \\
        \midrule
        \cite{Ladeveze2009}    & 2009 & --- & PBM, & \textsf{VirtuoseAPI}, & \textsf{Virtuose6D}$^\text{TM}$ \textsf{35-45}, & PRM PP methods (A-star, RDT/RRT);\\
        \cite{Ladeveze2010}    & 2010 &     & PP,  &              &  & guides parts along `following zones';\\
                               &      &     & CD.  &              &  & updates the path when user deviates;\\
        \midrule
        \cite{Zhen2009}& 2009 & \textsf{DPVAE} & PBM, & Device APIs, & \textsf{CyberGlove}$^\circledR$, & collaborative assembly environment;\\
        \cite{Wu2012}  & 2012 &                & CBM, & TCP/IP. & \textsf{CyberTouch}$^\circledR$, & supports data conversion from CAD;\\
                       &      &                & CD.  &  & \textsf{5DT FOB}$^\circledR$ \& DGs. & parallel rendering based on PC-cluster;\\
                       &      &                &      &  &  & `high-efficient' CD and HLA/RTI.\\
        \midrule
        \cite{Hu2010,Zhen2010}  & 2010 & \textsf{GCVAE}   & PBM, & Device APIs, & \textsf{CyberGlove}$^\circledR$, & collaborative assembly environment;\\
                                &      &                  & CBM, & TCP/IP. & \textsf{CyberTouch}$^\circledR$, & network grid‐based support platform;\\
                                &      &                  & CD.  &  & \textsf{5DT FOB}$^\circledR$ \& DGs. & supports large and complex scenes;\\
        \midrule
        \cite{Gutierrez2010} & 2010 & \textsf{IMA-VR} & CBM, &  \textsf{OpenHaptics}$^\circledR$, & \textsf{Phantom}$^\circledR$ \textsf{Omni}$^\circledR$, & multimodal assembly training system;\\
                             &      &     &  CD. & Device APIs. & CEIT \textsf{LHIfAM}, & cognitive and motor skills transfer;\\
                             &      &     &      &  & \textsf{PERCRO} \textsf{GRAB}. & spring-damper model for CD response.\\
        \midrule
        \cite{Tching2010,Tching2010a} & 2010 & --- & PBM, &  \textsf{VirtuoseAPI}. & \textsf{Virtuose6D}$^\text{TM}$ \textsf{35-45}, & VCG method for hybrid PBM+CBM;\\
                                      &      &     & CBM, &                        & & uses nonsmooth rigid body dynamics;\\
                                      &      &     & CD.  &                        & & uses virtual fixtures for DOF-limiting;\\
        \midrule
        \cite{Hassan2010}  & 2010 & \textsf{MAD} & PBM, & \textsf{OpenHaptics}$^\circledR$, & \textsf{Phantom}$^\circledR$ \textsf{Omni}$^\circledR$. & aircraft assembly and maintenance;\\
        \cite{Hassan2011,Hassan2011a}  & 2011 &  & CD.  & \textsf{OpenInventor}$^\circledR$ & & CACO path \& sequence optimization;\\
        \cite{Hassan2014}  & 2014 &              &      & \textsf{GLUT}. &  & active and passive haptic guidance.\\
        \midrule
        \cite{Webel2011,Webel2011a,Gavish2011}      & 2011 & --- & PBM, & Device APIs. & Haptic bracelet, & AR for maintenance \& assembly;\\
        \cite{Gutierrez2012}  & 2012 &     & CD.  &   & Generic webcam, & enables skills acquisition \& transfer;\\
        \cite{Webel2013}      & 2013 &     &      &   & Generic tablet. & AVA and vibrotactile feedback; \\
        \cite{Gavish2015}     & 2015 &     &      &   &  & low-cost AR training platform.\\
        \midrule
        \cite{Xia2011} & 2011 & \textsf{HVAS} & PBM, & \textsf{OpenHaptics}$^\circledR$. & \textsf{Phantom}$^\circledR$ \textsf{Premium}$^\circledR$, & combined PBM+CBM for assembly;\\
                       &      &               & CBM, & \textsf{WTK} Toolkit, & Unspecified SG$^\ast$. & automatic data transfer from CAD;\\
                       &      &               & CD.  & \textsf{PhysX}$^\text{TM}$ SDK. &  & hierarchical constraint data model.\\
        \midrule
        \cite{Xia2012} & 2012 & \textsf{HITsphere} & PBM, & \textsf{OpenHaptics}$^\circledR$. & \textsf{Phantom}$^\circledR$ \textsf{Premium}$^\circledR$, & motion simulator for free walking;\\
                       &      &  & CBM, & \textsf{WTK} Toolkit, & \textsf{Cybersphere} System, & combined PBM+CBM for assembly;\\
                       &      &  & CD.  & \textsf{PhysX}$^\text{TM}$ SDK, & Unspecified SG$^\ast$. & automatic data transfer from CAD;\\
                       &      &  &      & \textsf{TechViz XL}. &  & hierarchical constraint data model.\\
        \midrule
        \cite{Murakami2013} & 2013 & \textsf{Poster} & PBM, & \textsf{PTAMM}, \textsf{ODE}, & \textsf{HapticGEAR}, & wearable backpack-type haptic device;\\
                            &      &                 & CD.  &  \textsf{ARToolKit}.  & Unspecified HMD, & markerless AR with large workspace.\\
        \midrule
        \cite{Olsson2013}   & 2013 & \textsf{Snap-to-Fit} & PBM, & \textsf{H3DAPI}, & Unspecified Device, & point-to-point attraction force model;\\
                            &      &                      & CD.  &   &  & applied to facial surgery \& archaeology.\\
        \midrule
        \cite{Behandish2014,Behandish2014a} & 2014 & --- & PBM, & \textsf{OpenHaptics}$^\circledR$, & \textsf{Phantom}$^\circledR$ \textsf{Omni}$^\circledR$, & unified CD and geometric guidance;\\
        \cite{Behandish2015,Behandish2015a} & 2015 &     & CD,  & \textsf{VirtuoseAPI}, & \textsf{Virtuose6D}$^\text{TM}$ \textsf{35-45}. & automatic GE for arbitrary geometry;\\
        \cite{Behandish2016}                & 2016 &     & GE.  & \textsf{OpenGL}, \textsf{Havoc3D}. &  & does not depend on CAD constraints;\\
                                            &      &     &      & \textsf{GLUT}, \textsf{Win32} API.  &  & does not scale with syntactic size;\\
        \bottomrule
    \end{tabular}
    \begin{tablenotes}
        \footnotesize
        \item \textbf{Abbreviations used for generic terms:} physically-based modeling (PBM), constraint-based modeling (CBM), collision detection (CD), virtual environment (VE), virtual reality (VR, augmented reality (AR), computer-aided design (CAD), application programmer's interface (API), software development kit (SDK), transmission control protocol (TCP), user datagram protocol (UDP), internet protocol (IP), optical see-through (OST), head-mounted display (HMD), boundary representation (B-rep), shutter glasses (SG), stereo glasses (SG$^\ast$), projection table (PT), task completion time (TCT), electromyography (EMG), motion timeline (MTL), physics simulation engine (PSE), data glove (DG), stereo projector (SP), self-efficacy (SE), perceived virtual environment (PVE), axis-aligned bounding box (AABB), optical tracker (OT), probabilistic roadmap (PRM), path planning (PP), rapidly-growing deterministic trees (RDT), and rapidly-exploring random trees (RRT), high level architecture (HLA), runtime infrastructure (RTI), virtual constraint guidance (VCG), degrees of freedom (DOF), combined ant colony optimization (CACO), adaptive visual aids (AVA), geometric energies (GE).
        \item \textbf{Abbreviations used for brand names:} \textsf{virtual environment for design for assembly (VEDA), general haptic open software toolkit (GHOST), Silicon Graphics, Inc. (SGI), virtual assembly design environment (VADE), Flock of Birds (FOB), haptic integrated dis/reassembly analysis (HIDRA), open graphics library (OpenGL), Voronoi clip (V-Clip), speedy walking via improved feature testing (SWIFT), realidad virtual para el estudio de mantenibilidad en sistemas aeron\`{a}uticos (REVIMA), large haptic interface for aeronautic maintainability (LHIfAM), multimodal immersive virtual assembly system (MIVAS), cave automatic virtual environment (CAVE), system for haptic assembly and realistic prototyping (SHARP), Voxmap PointShell (VPS), haptic assembly, manufacturing, and machining system (HAMS), haptic assembly and manufacturing system (HAMS), visualization toolkit (VTK), Microsoft Foundation Class (MFC), perceptual robotics (PERCRO), haptic assembly simulator (HAS), collaborative haptic assembly simulator (CHAS), open/object dynamics engine (ODE), haptically enabled interactive and immersive virtual reality (HIIVR), computer haptics and active interfaces 3D (CHAI3D), virtual environment for the design and assembly planning (VEDAP), virtual human toolkit (VHT), interactive physics pack (IPP), National Aeronautics and Space Administration (NASA), synchronized position hold, engage, reorient, experimental satellites (SPHERES), distributed parallel virtual assembly environment (DPVAE), grid‐enabled collaborative virtual assembly environment (GCVAE), industrial maintenance and assembly with virtual reality (IMA-VR), maintenance assembly/disassembly (MAD), graphics library utility toolkit (GLUT), haptic-based virtual assembly system (HVAS), world toolkit (WTK), parallel tracking and multiple mapping (PTAMM).}
        \item {}{$\times2$} means two-handed haptic device or a pair of one-handed devices.
    \end{tablenotes}
\end{threeparttable}
\end{adjustbox}
\end{table*}

Vance et al. at the Iowa State University developed a series of VA tools with and without haptic support: Johnson and Vance developed \textsf{VEGAS} \cite{Johnson2001}, an assembly simulator that used \textsf{Boeing Corp.}'s \textsf{Voxmap PointShell (VPS)}$^\text{TM}$ library \cite{Wan2003,McNeely2005a,McNeely2005,McNeely2006} for CD between high-polygon parts without PBM for haptic feedback. Kim and Vance investigated different CD and part behavior algorithms \cite{Kim2003,Kim2004} and modified \textsf{VEGAS} to include PBM but no haptics. Kim and Vance also developed \textsf{NHE} \cite{Kim2004a} to facilitate collaborative assembly over the web using a combination of peer-to-peer and client-server models. Howard and Vance \cite{Howard2007} developed a prototype desktop system for haptic assembly using PBM. From the same group, Seth et al. later developed \textsf{SHARP} \cite{Seth2005,Seth2006,Seth2008,Seth2007,Seth2010} which expanded the VA functionality to include dual-handed haptics, swept volume representation, subassembly modeling, and more realistic part behavior via PBM. The earlier versions of \textsf{SHARP} \cite{Seth2005,Seth2006,Seth2008} used \textsf{VPS}$^\text{TM}$ for approximate CD between a `voxmap' representation of an stationary part and a `pointshell' representation of the moving part, which was inadequate for low-clearance assembly. The subsequent versions of \textsf{SHARP} \cite{Seth2007,Seth2010} implemented a combination of PBM and CBM using the automatic geometric constraints (AGC) method \cite{Vance2011}; it used \textsf{D-Cubed}'s \textsf{CDM} module for exact CD between original CAD B-rep data and \textsf{D-Cubed}'s \textsf{DCM} module for constraint management. Faas and Vance \cite{Faas2011} proposed a hybrid method to tie the B-rep and voxel representations for simultaneous collision response and constraint-based guidance. Dual \textsf{Phantom}$^\circledR$ \textsf{Omni}$^\circledR$ devices (6 DOF input, 3 DOF output) were used.

Cheng-Jun et al. at the Qingdao Technological University proposed the use of dynamically constructed oriented bounding box (OBB) tree-based CD \cite{Gottschalk1996} for haptic assembly \cite{Cheng-Jun2007}. They also proposed a `increment-along-constraint' (IAC) method \cite{Cheng-Jun2010} to solve the separation problem between the haptic proxy and the parts under constrained motion. A \textsf{Phantom}$^\circledR$ \textsf{Desktop}$^\circledR$ device (6 DOF input, 3 DOF output) was used, for which the specialized OBB and IAC algorithms were implemented.

Lim et al. at the Heriot-Watt University developed a system called \textsf{HAM(M)S} \cite{Lim2006,Lim2006a,Lim2007a,Lim2010,Lim2007,Badillo2013,Badillo2014} as a testbed to investigate and measure user interactions and response while performing various engineering tasks in a haptic-enabled VE including assembly.%
\footnote{The system was called \textsf{haptic assembly, manufacturing, and machining system (HAMMS)} in the earlier publications \cite{Lim2006,Badillo2013}. The authors chose to simplify the system's name to \textsf{haptic assembly and manufacturing system (HAMS)} in a subsequent paper \cite{Badillo2014}.}
Different physics simulation engines were utilized; namely \textsf{Ageia PhysX}$^\text{TM}$ SDK \cite{Chan2009} was used in the earlier versions \cite{Lim2007a} and \textsf{Bullet Physics} SDK \cite{Sagardia2014} was added into the later versions \cite{Badillo2013,Badillo2014}, whose pros and cons were evaluated and compared in \cite{Badillo2014a,Badillo2014b}. They conducted peg-in-hole assembly experiments (using insertion routines with and without chamfers) on both real and virtual setups for comparison in terms of task completion times (TCT) \cite{Lim2007} and motor control measured via muscle Electromyography (EMG) \cite{Lim2010}. Further investigation of the user-object interaction was made with the objective of assembly plan generation by analyzing chronocycle-graph motion timelines (MTL) and ``therblig'' units \cite{Sung2008} for both peg-in-hole examples and a more realistic pump assembly \cite{Ritchie2009,Ritchie2008b,Ritchie2008c}. \textsf{Phantom}$^\circledR$ \textsf{Omni}$^\circledR$ and \textsf{Desktop}$^\circledR$ devices (6 DOF input, 3 DOF output) were used interchangeably. The group also developed a cable harness VR system called \textsf{COSTAR} \cite{Sung2009,Sung2008,Sung2009a} (without haptic feedback) for user-logging in order to analyze the design process, capture design knowledge, and produce assembly plans.

The Fundaci\'{o}n LABEIN research team (Iglesias et al.) at Tecnalia developed the distributed haptic assembly system \textsf{HAS} \cite{Iglesias2006a} using their own geometric modeler \textsf{DATum} for creation of 3D virtual scenes from CAD models. An assembly simulator was developed based on \textsf{RAPID} \cite{Gottschalk1996} for CD and simple semantics for constraint detection and enforcement; namely, along co-axial axes and co-planar planes. However, it did not include real-time physics-based simulation. The collaborative extension called \textsf{CHAS} \cite{Iglesias2006,Iglesias2007,Iglesias2008} using a peer-to-peer architecture and specific consistency maintenance schemes was developed later and tested on an aeronautical assembly test-case. \textsf{Phantom}$^\circledR$ \textsf{Omni}$^\circledR$ and \textsf{Premium}$^\circledR$ devices (6 DOF input, 3$-$6 DOF output) and a PERCO \textsf{GRAB} device providing two points of contact (4 DOF input, 4 DOF output) were used.

Christiand et al. at the Gyeongsang National University developed an assembly simulation system \cite{Christiand2011,Christiand2008,Christiand2009} with haptic guidance along an optimized path. The sequence identification and path planning were carried out offline using an optimization genetic algorithm (GA) that took into account the part geometries and gripper data. Once a sequence was identified and a path was generated for each part in the sequence, a potential field method was used in real-time to combine repulsive and attractive forces to avoid obstacles and guide each part to its final position along the known path. The implementation was limited to the assembly of 2D polygonal parts whose collisions were detected using an axis-aligned bounding box (AABB) method. The experiments concluded better result with haptic guidance compared to those without haptics, both in terms of assembly time and travel distance. Hassan and Yoon from the same group developed the assembly and disassembly maintenance system called \textsf{MAD} \cite{Hassan2010}. The system imported CAD files from \textsf{CATIA}$^\circledR$ using \textsf{OpenInventor}$^\circledR$ API and performs sequence and path planning using parallel GA-based planners. Given the initial and final configurations, the GA minimizes the number of required gripper exchanges and orientation changes to reduce the assembly time according to the findings in \cite{Pan2006}. To accommodate the geometric complexity and high-DOF that led to the failure of GA, Hassan and Yoon later developed a two-stage combined ant colony optimization (CACO) algorithm \cite{Hassan2011a}, which performed sequence optimization in the first stage followed by path planning in the second stage. The CACO algorithm was introduced to \textsf{MAD} \cite{Hassan2011} to investigate the effects of active and passive haptic guidance to improve the user performance. A \textsf{Phantom}$^\circledR$ \textsf{Omni}$^\circledR$ device (6 DOF input, 3 DOF output) was used for all experiments.

The KAEMaRT research group (Cugini et al. and Bordegoni et al.) at the Politecnico di Milano evaluated haptic-assisted manual assembly in a mixed reality environment for the case study of grabbing, holding, and positioning pairs of mechanical components relative to each other \cite{Bordegoni2009}. A \textsf{Virtuose6D}$^\text{TM}$ \textsf{35-45} (6 DOF input, 6 DOF output) and a \textsf{Nintendo}$^\text{TM}$ \textsf{WiiRemote}$^\text{TM}$ were used together in a two-handed setup. They used a set of heuristic criteria for assessing the quality of the application and identifying usability problems to be fixed in future studies.

Ladeveze et al. at the Universit\'{e} de Toulouse developed a system for haptic assembly and disassembly task assistance \cite{Ladeveze2010,Ladeveze2009} using probabilistic roadmap (PRM) path planners---namely, tools such as the A-star algorithm, rapidly-growing deterministic trees (RDT), and rapidly-exploring random trees (RRT) \cite{Aguinaga2007}. The planner first identified a collision free path in the 6D configuration space of a rigid part. The haptic control loop then used that information to guide the user into and along a so-called `following zone' formed around the free path, which was discretized for simplifying and stabilizing force and torque computation. A \textsf{Virtuose6D}$^\text{TM}$ \textsf{35-45} device (6 DOF input, 6 DOF output) was used.

Wu et al. at the Shanghai Jiao Tong University developed a grid-based VA server called \textsf{GVAS} \cite{Zhen2010} for large and complex (e.g., automotive and ship) assemblies, which used parallel computing and network resources for demanding VA computations such as model rendering, image processing (i.e., fusion), and CD. To fulfill security requirements, product data was managed independently using the concept of role-based access control (RBAC). The group later developed two systems for automobile VA; namely: 1) a grid-based collbaorative VA environment called \textsf{GCVAE} \cite{Hu2010}, which comprised of a grid-based support platform, a service-based parallel rendering framework, and a multi-user collaborative VA environment; and 2) a distributed parallel VA environment called \textsf{DPVAE} \cite{Zhen2009,Wu2012}, which used high-level architecture and runtime infrastructure (HLA/RTI) event synchronization mechanisms. The systems made extensive use of parallel processing to enable complex assembly scenarios with intensive rendering and CD requirements at multiple levels of detail (LOD). \textsf{CyberTouch}$^\circledR$ haptic gloves (with small vibrotactile feedback) were used.

Tching et al. at the IRISA--Bunraku developed the virtual constraint guidance (VCG) method \cite{Tching2010,Tching2010a} for haptic guidance. The method decomposes a task into 1) a guiding step which used virtual fixtures \cite{Rosenberg1993} to guide the objects into position; and 2) a functional step which used kinematic constraints via mechanical joints to restrict the DOF for insertion. The idea is to use both PBM-based exploration of the VE (using nonsmooth dynamics \cite{Tching2008}) and CBM-based execution of fine insertion (using virtual fixtures \cite{Rosenberg1993}) while CD is locally deactivated. The method has proved very effective for peg-in-hole test-cases with simple contact constraints (i.e., lower pairs) whose virtual fixture abstractions and mechanical joint equivalents are obvious. The constraints were extracted from CAD mating pair semantics in a preprocessing step. A \textsf{Virtuose6D}$^\text{TM}$ \textsf{35-45} (6 DOF input, 6 DOF output) was used.

Xia et al. at the University of Porto developed a multithreaded haptic assembly system called \textsf{HVAS} \cite{Xia2011}. The system had an automatic data integration interface to transfer geometry, topology, assembly, and physics information from CAD systems (e.g., \textsf{Pro/ENGINEER}$^\circledR$ and \textsf{SolidWorks}$^\circledR$) to the VE. A hierarchical constraint-based data model and scene-graph structure was designed to construct the VA environment. The system employed a combined PBM and CBM approach for haptic guidance; namely, a spring-damper model for collision response and dynamic simulation when parts penetrate into each other, and a geometric guidance model to generate attractive and repulsive forces to guide the mating parts that are close to each other. The architecture was embedded into another system called \textsf{HITsphere} \cite{Xia2012} for VA training which, unlike the traditional desktop or \textsf{CAVE}-like systems, also enabled natural human walking motion---similar to the \textsf{Cybersphere} system \cite{Fernandes2003}---using a low-cost motion simulator. Haptic feedback was enabled by a \textsf{Phantom}$^\circledR$ \textsf{Premium}$^\circledR$ device (6 DOF input, 6 DOF output).

The G-SCOP group (No{\"e}l et al.) at the Grenoble Institute of Technology and Iacob et al. at the University Politehnica of Bucharest developed a collaborative VE called \textsf{CVE} \cite{Trakunsaranakom2014,Popescu2015} for design and assembly activities, which contained several modules---such as viewer, recorder, editor, \textsf{object dynamics engine (ODE)}, analysis, and haptic models---integrated together for different sub-behaviors. The system performed automatic constraint detection using a contact identification process between `functional surfaces'---restricted to planar, cylindrical, spherical, and conical surfaces---between CAD models (exported in STEP format) \cite{Iacob2007,Iacob2008,Iacob2011,Iacob2013,Iacob2012}. The system was evaluated for usability, efficiency, user experience, and feedback quality. A stereoscopic display and a \textsf{Virtuose6D}$^\text{TM}$ \textsf{35-45} device (6 DOF input, 6 DOF output) was used for haptic feedback.

Table \ref{tab_VAlit} presents a more inclusive list (including studies and systems not described above) and a summary of their key features.

\bibliographystyle{plain}
\singlespacing{\footnotesize \bibliography{CDL-TR-16-04}}
\end{document}